\documentclass[traditabstract]{aa} 

\usepackage[colorlinks=true, linkcolor=blue, citecolor=blue, urlcolor=blue]{hyperref}
\usepackage{amsmath}
\usepackage{graphicx}
\usepackage{txfonts}
\usepackage{amssymb}
\usepackage[]{natbib}
\usepackage{mathrsfs}
\usepackage{caption}
\usepackage{float}
\usepackage{afterpage}
\usepackage{amstext}

\newcommand{\wasp}[0]{\mbox{WASP-33}}
\newcommand{\waspb}[0]{\mbox{WASP-33~b}}
\newcommand{\waspa}[0]{\mbox{WASP-33A}}

\newcommand{\myrev}[1]{{\bf}}

\begin{document}

   \title{An optical transmission spectrum of the ultra-hot Jupiter WASP-33b}
   \subtitle{First indication of AlO in an exoplanet}
 \author{C. von Essen\inst{1}
          \and
          M. Mallonn\inst{2}
          \and
          L. Welbanks\inst{3}
          \and
          N. Madhusudhan\inst{3}
          \and
          A. Pinhas\inst{3}
          \and
          H. Bouy\inst{4}
          \and
          P. Weis Hansen\inst{1}
          }

   \institute{Stellar Astrophysics Centre, Department of Physics and Astronomy, Aarhus University, Ny Munkegade 120, DK-8000 Aarhus C, Denmark\\
         \email{cessen@phys.au.dk}
         \and
             Leibniz-Institut f\"{u}r Astrophysik Potsdam (AIP), An der Sternwarte 16, D-14482 Potsdam, Germany
         \and
             Institute of Astronomy, University of Cambridge, Madingley Road, Cambridge CB3 0HA, UK
        \and
             Laboratoire d'astrophysique de Bordeaux, Univ. Bordeaux, CNRS, B18N, allée Geoffroy Saint-Hilaire, 33615 Pessac, France.
             }
   \date{Received 12/07/2018; accepted 10/10/2018}
   
\abstract{There has been increasing progress toward detailed
  characterization of exoplanetary atmospheres, in both observations
  and theoretical methods.  Improvements in observational facilities
  and data reduction and analysis techniques are enabling increasingly
  higher quality spectra, especially from ground-based facilities. The
  high data quality also necessitates concomitant improvements in
  models required to interpret such data. In particular, the detection
  of trace species such as metal oxides has been
  challenging. Extremely irradiated exoplanets ($\sim$3000 K) are
  expected to show oxides with strong absorption signals in the
  optical. However, there are only a few hot Jupiters where such
  signatures have been reported. Here we aim to characterize the
  atmosphere of the ultra-hot Jupiter \waspb\ using two primary
  transits taken 18 orbits apart. Our atmospheric retrieval, performed
  on the combined data sets, provides initial constraints on the
  atmospheric composition of \waspb. We report a possible indication
  of aluminum oxide (AlO) at 3.3-$\sigma$ significance. The data were
  obtained with the long slit OSIRIS spectrograph mounted at the
  10-meter Gran Telescopio Canarias. We cleaned the brightness
  variations from the light curves produced by stellar pulsations, and
  we determined the wavelength-dependent variability of the planetary
  radius caused by the atmospheric absorption of stellar light. A
  simultaneous fit to the two transit light curves allowed us to
  refine the transit parameters, and the common wavelength coverage
  between the two transits served to contrast our results. Future
  observations with HST as well as other large ground-based facilities
  will be able to further constrain the atmospheric chemical
  composition of the planet.}

          
   \maketitle

\section{Introduction}

Transiting exoplanets offer a unique opportunity to determine the
physical and chemical properties of their atmospheres. In particular,
the transmission spectra observed when the planet transits in front of
its host star can reveal the composition and extent of the atmosphere
at the day-night terminator region of the planet. Here, the extinction
of the stellar photons travelling through the planetary atmosphere
along the line of sight is imprinted on the observed stellar spectrum
during the transit. Time-domain differential spectroscopy, between in
and out of transit, reveals the absorption spectrum of the planetary
atmosphere, also known as a "transmission" spectrum. This technique
and its variants have been used to detect several atomic and molecular
species in exoplanetary atmospheres; for example sodium
\cite{Charbonneau2002,Snellen2008,Redfield2008}, H$_2$O and CO
\cite{Snellen2010,Deming2013,Kreidberg2014}, and clouds and hazes
\cite{Bean2010,Pont2013,Kreidberg2014a}.

To characterize the chemical compositions of exoplanetary atmospheres
two approaches of transit spectroscopy have been successfully
used. High resolution transmission spectroscopy is ideally suited to
resolve individual absorption lines of the chemical species in the
atmosphere, in other words, via the imprint of the planetary lines on
the stellar spectrum
\citep[e.g.,][]{Snellen2008,Wyttenbach2015,Khalafinejad2017}. This
technique first led to the detection of sodium in the atmosphere of
\mbox{HD 209458b} from high-precision spectro-photometric observations
\citep{Charbonneau2002}. At resolutions of \mbox{R$\sim$100 000}, the
absorption line cores of individual chemical species such as sodium
can be spectroscopically resolved, as well as the orbital motion, the
diurnal rotation, and in some extreme cases exo-atmospheric wind
speeds \citep{Snellen2010,Louden2015}. On the other hand, the use of
low resolution spectra and broadband photometry allows investigations
of broad spectral features over a wide wavelength range, such as the
presence of clouds, hazes or Rayleigh scattering, along with strong
atomic and molecular absorbers
\citep[e.g.,][]{Sing2016,Mallonn2016b,vonEssen2017,Mallonn2017,Nikolov2018}.

In tandem with observational advancements, there have also been key
developments in atmospheric modeling of exoplanets. Several studies
predict key atomic and molecular species to be present in exoplanetary
atmospheres depending on the macroscopic conditions such the
equilibrium temperature, gravity, and metallicity \citep[see
  e.g.,][]{Seager2000, Hubbard2001, Fortney2010, madhu2016}. Early
theoretical modeling suggested Hot Jupiters with equilibrium
temperatures above 2500 K to be similar to M dwarfs containing gaseous
oxides such as TiO and VO in cloud-free atmospheres
\citep{Hubeny2003,Fortney2008}. The strong opacity of TiO/VO at
optical wavelengths could result in thermal inversions, that is,
rising temperatures with higher altitudes \citep{Burrows2008}. The
first observational evidence for such inversion layers were published
for a much cooler Hot Jupiter \citep{Knutson2008}. However, this and
several other measurements were revised subsequently \citep[see
  e.g.,][]{Diamond-Lowe2014,Hansen2014,Evans2015}. In parallel,
several mechanisms were investigated explaining why gaseous TiO might
not play a role in the physics of the high altitude layers, including
the gravitational settling to colder atmospheric layers were the
molecules condense out as well as the carbon to oxygen ratio
\citep{Spiegel2009,Parmentier2013,madhu2012}.

Recently, promising indications for the presence of TiO have been
published for two of the hottest gas giants known. In the terminator
region of the $T_{\mathrm{eq}} = 2400$~K WASP-121b, \cite{Evans2016}
found hints of additional opacity at optical wavelengths, indicative
of TiO. Shortly thereafter, \cite{Evans2017} found stronger evidence
for TiO and a thermal inversion in the day-side of the same
planet. Transmission spectroscopy of \mbox{WASP-19b} ($T_{\mathrm{eq}}
= 2100$~K) resolved optical absorption bands from TiO
\citep{Sedaghati2017}. However, measurements for the similarly hot gas
giants \mbox{WASP-12b} ($T_{\mathrm{eq}} = 2600$~K) showed no signs of
TiO absorption at the terminator, making it more likely that these
elements have been removed from these parts of the planetary
atmosphere \citep{Sing2013}. Another difference between these three
very hot gas giants is the level to which the transmission spectra are
dominated by clouds and hazes. While it is substantial enough for
\mbox{WASP-12b} to cause a featureless optical spectrum
\citep{Sing2013}, the atmospheres of WASP-19b and WASP-121b seem to be
less affected, showing absorption features. The formation of
condensate clouds in ultra-hot Jupiter atmospheres was investigated by
\cite{Wakeford2017}.

The TiO signals in WASP-19b and WASP-121b have so far been the
strongest evidence for the presence of the debated gaseous TiO,
indicating a dependence of temperature. One of the hottest planets
after KELT-9b is \waspb\ with measured brightness temperature at its
dayside of \mbox{T$_{\rm B}$ = 3398 $\pm$ 302 K}
\citep{vonEssen2015}. Very recently, \cite{Nugroho2017} found
significant signal of TiO in its dayside including indications for a
thermal inversion. The existence of this stratosphere in its dayside
was also suggested by \cite{Haynes2015}. An investigation of the phase
curve of \waspb\ by \cite{Zhang2018} described the planets heat
recirculation as similar to cooler Hot Jupiters.

In this work we present the search for oxides and clouds and hazes at
the terminator region of the atmosphere of
\waspb\ \citep{CollierCameron2010} by transmission spectroscopy in the
optical. We exploit two transit observations taken 18 orbits apart
with the 10 meter Gran Telescopio Canarias.  The exoplanet orbits an
A-type star every $\sim$1.22 days, making it one of the strongest
irradiated planets known till date \citep{Smith2011,vonEssen2015}. The
host is a $\delta$ Scuti star, with pulsation amplitudes in the
milli-magnitude regime and frequencies on the order of a few hours
\citep{Herrero2011,vonEssen2014}. \cite{Lehmann2015} carried out a
spectroscopic follow-up of the host star, characterizing the mass of
the planet to be around 2.1 M$_J$. \waspb\ shows an unusually large
radius ($\sim$1.6 R$_J$), making it inflated
\citep{CollierCameron2010} and ideal for transmission spectroscopy
studies. Along a 2.5-year photometric follow up we have characterized
the pulsation spectrum of the host star \citep{vonEssen2014}. From
this analysis we found eight statistically significant pulsation
frequencies, most of them well contained within a typical ground-based
observation. Their proper representation is fundamental to correctly
determine the transmission spectrum of \waspb. In this work,
Section~\ref{Obs_Data} presents the observations and the data
reduction, Section~\ref{Fitting} shows a detailed description of the
model parameters and fitting procedures, Section~\ref{TS} shows our
results on the transmission spectrum of \wasp and constraints on its
atmospheric properties, while Section~\ref{Concl} contains our final
remarks.

\section{Observations and data reduction}
\label{Obs_Data}

\subsection{Instrumental setup}
 
On the nights of August 7th (henceforth observing run 1, OR1) and
August 30th (observing run 2, OR2), 2014, we simultaneously observed
\wasp\ \mbox{(V = 8.14)} and the star \mbox{BD+36 488} \mbox{(V =
  9.37)} during primary transit. For our observations we used the
longslit spectroscopic mode of OSIRIS, mounted at the 10-meter Gran
Telescopio Canarias, Spain. The observations were carried out in its
standard observing mode (binning 2$\times$2). To avoid saturation of
the target we observed with the telescope slightly defocused. To
minimize flux losses \citep[e.g.,][]{Sing2012} the slit was opened to
10 arcsec.

\begin{figure}[ht!]
  \centering
  \includegraphics[width=.5\textwidth]{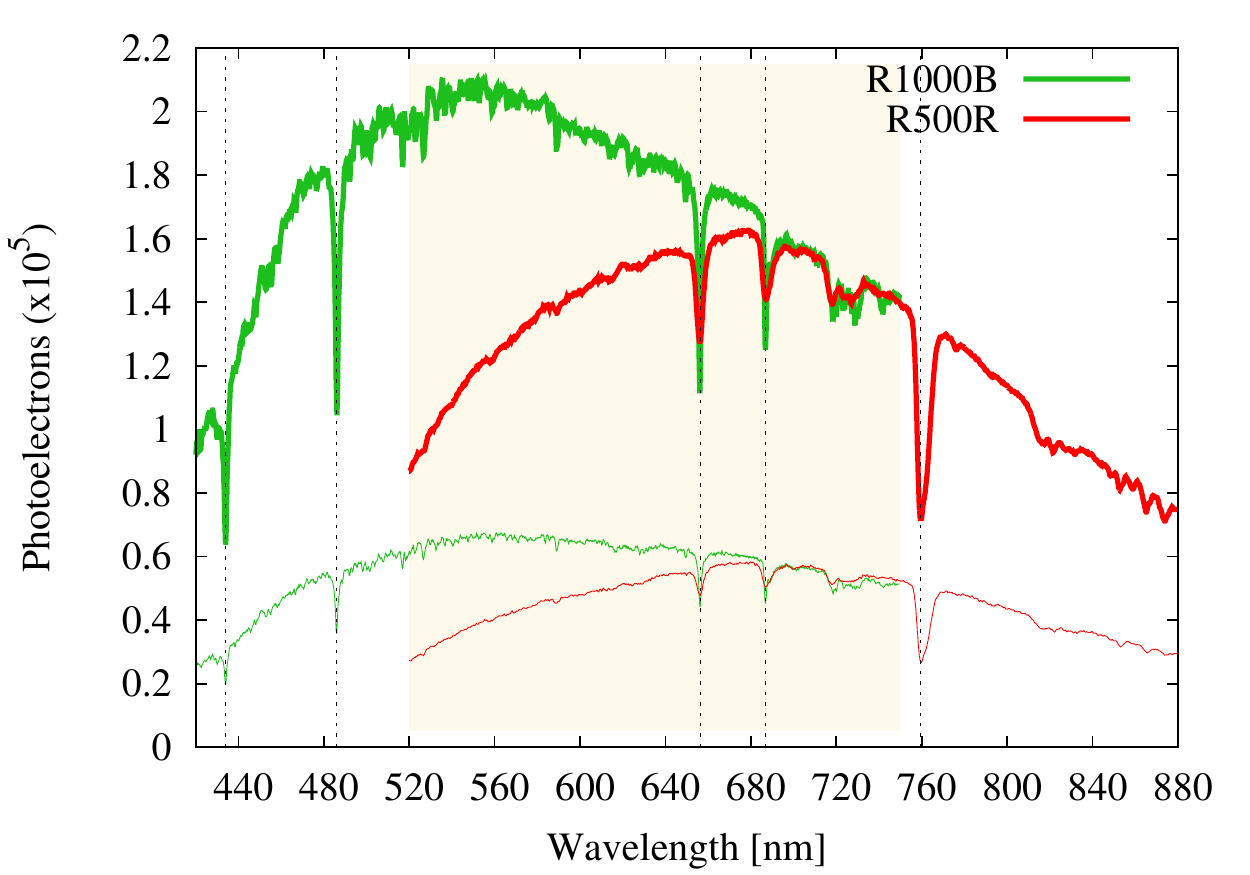}
  \caption{\label{fig:wav_vs_flux} Extracted spectra in photoelectrons
    for WASP-33 (continuous thick lines) and \mbox{BD+36 488}
    (continuous thin lines) for the R1000B and R500R grisms (green and
    red colors, respectively). Vertical dashed black lines indicate
    the atmospheric and stellar lines used to align the spectra. The
    yellow shaded region highlights the wavelength coverage of the
    data sets used to construct the white light curves (see
    Section~\ref{WLC}).}
\end{figure}

During OR1 we observed \wasp\ using the R1000B grism (spectral
resolution, \mbox{R = 1018}), which allows for a wavelength coverage
between 363 and 750 nm. We collected 587 spectra along a total of 4.24
hours, starting at 1:31 UT and finishing at 5:45 UT. Our observations
cover the primary transit fully, plus about one our before and one
hour after transit, respectively. The exposure time was set to one
second. Due to readout time the final cadence of the data was 25
seconds. For OR1 we integrated fluxes between 420 and 750 nm, avoiding
the wavelengths where the signal was too low. During OR2 we observed
\wasp\ for 4.4 hours using the R500R grism \mbox{(R = 587)}, covering
the wavelengths between 480 and 1000 nm. The exposure time was set to
0.4 seconds. Using a faster readout mode, the final cadence was 19
seconds. The observations started at 0:11 UT and ended at 4:35 UT,
which corresponds to full transit coverage plus an hour before and
after transit, respectively. We considered fluxes between 520 and 880
nm, and neglected the wavelength region around telluric oxygen
\mbox{(760 $\pm$ 10 nm)}. The considered wavelength coverage for each
one of the grisms, along with the instrumental fluxes of \wasp\ and
\mbox{BD+36 488}, can be seen in Figure~\ref{fig:wav_vs_flux}.

\subsection{Spectral extraction and determination of environmental quantities}

\begin{figure}[ht!]
  \centering
  \includegraphics[width=.5\textwidth]{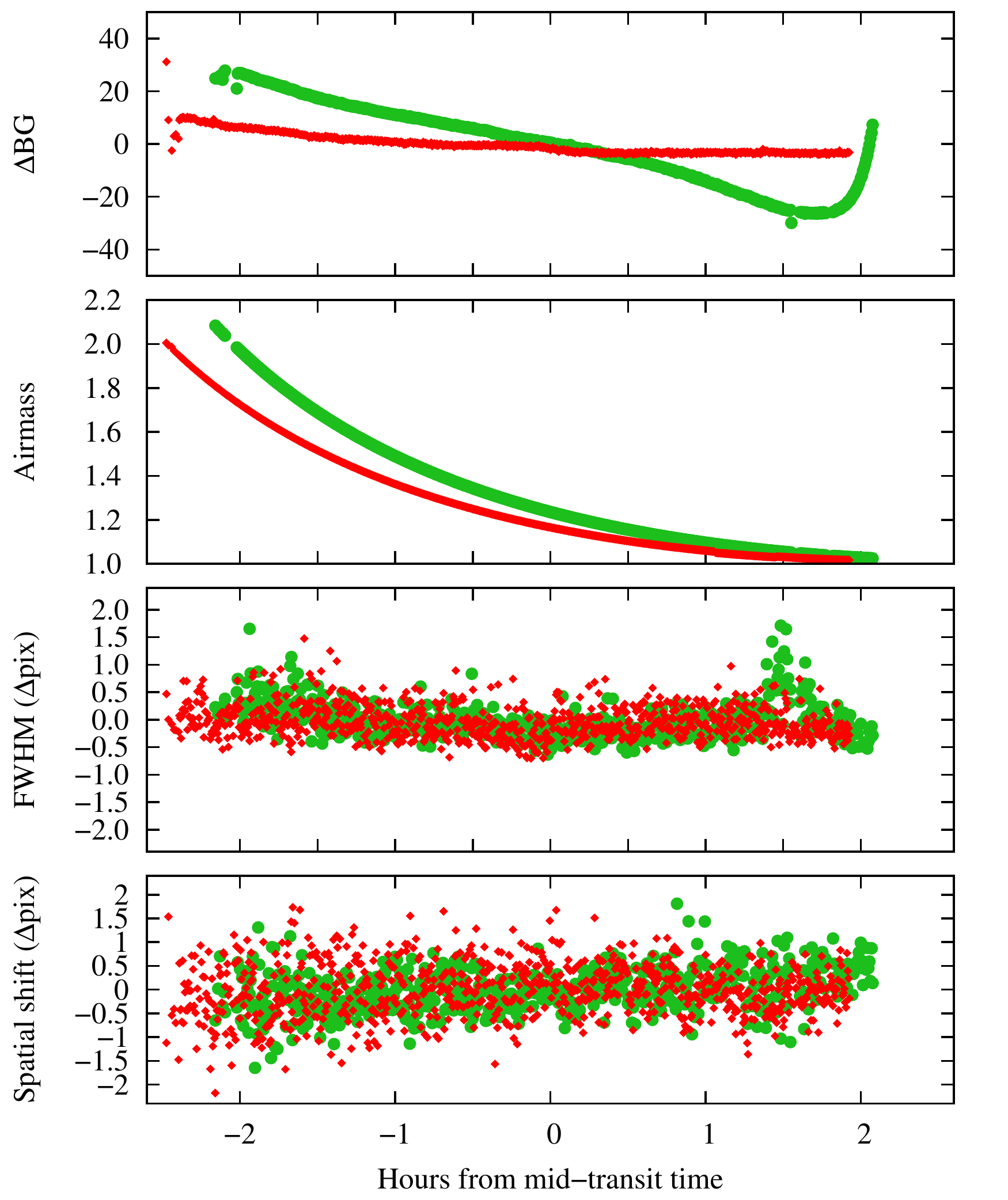}
  \caption{\label{fig:environmental} From top to bottom,
    time-dependent changes of background counts (BG), airmass, FWHM,
    and spatial shifts. Both data sets have been shifted to their
    respective mid-transit times, and are shown in hours. Large green
    circles correspond to the R1000B (OR1) data, while small red
    diamonds correspond to R500R (OR2).}
\end{figure}

Before extracting the spectra, we began our analysis by computing the
changes in airmass, seeing, centroid positions and background
analyzing each stellar spectra. Airmass values were extracted from the
header of the images. We fitted a Gaussian function to several cuts
along the two-dimensional spectra. The fitting parameters are the
means, $\mu$ and standard deviations, $\sigma$. The one-on-one
differences of the means between the frame with the lowest airmass
(i.e., the reference frame) and all the remaining frames were averaged
and used to estimate the spatial shifts. The changes of full-width at
half-maximum (FWHM) were determined from \mbox{FWHM =
  2$\sqrt{2ln(2)}<\sigma>$}, where $<\sigma>$ is the average of the
standard deviations. Background values were computed averaging the
number of counts in predefined regions left and right from \wasp's
spectra, along all wavelengths. Figure~\ref{fig:environmental} shows
these quantities as a function of time. In the figure, time is given
in hours from mid-transit time. Both nights are photometrically
stable, and the spatial shifts are well contained within two binned
pixels. Due to the observed spatial stability, we did not flat-field
the spectra. This would only add an unnecessary source of noise. To
test this, we produced light curves with and without flatfielding the
spectra, finding no significant difference with respect to the
photometric precision of the derived spectro-photometric light
curves. The light curves produced from flatfielded data showed only
slightly larger photometric scatter.

To avoid saturation of the target, the telescope was scarcely
defocused. In consequence, we can not report averaged seeing values
from our data. Typical seeing on Roque de los Muchachos observatory is
about or below 1 arcsec. During OR1 the FWHM was $\sim$2.7 arcsec,
while for OR2 it was around 2.8 arcsec, thus we do not expect
significant flux loses in the 10$''$ slit. We carried out the
wavelength calibration in the usual way, using the {\it identify} IRAF
task. The RMS of the pixel-to-wavelength conversion was, in both runs,
lower than 0.05 arccec/pixel, well contained within the spectral
resolution. Since the exposure time was low in both observing nights,
we did not carry out cosmic ray removal.

Before extracting stellar fluxes, we carried out the background
subtraction with IRAF's task {\it apall}, fitting the sky spectrum
with a constant offset along predefined background regions at least 10
FWHM away from the \wasp\ spectra in the spatial direction spectra. To
test the impact of the aperture size into the accuracy of the
spectro-photometric light curves, we extracted the one-dimensional
spectra using IRAF's {\it apall} task with apertures of between 1 and
15 arcseconds in steps of 1 arcsecond. For this exercise, photometric
light curves were produced integrating fluxes between 520 and 750 nm
for both OR1 and OR2. We then computed the standard deviation of each
light curve taking into account only the off-transit data points, and
chose the final apertures that minimized this standard deviation. As
an independent test we also fitted a high order time-dependent
polynomial to the data, that accounted for the transit shape and the
pulsations of the host star. We then computed the standard deviation
of the residuals, obtained subtracting this best-fit nonphysical
polynomial to the photometric data sets. In both cases, and for both
transits, the aperture minimizing the previously mentioned standard
deviations was found to be 6 arcseconds. It is worth to mention that
large apertures minimizing the standard deviation of the data have
been observed before in spectrophotometric transit observations with
OSIRIS \citep{Mallonn2015,Mackebrandt2017,Chen2017}.

Throughout this work, time stamps are converted from the Julian Dates
extracted from the header of the images to Barycentric Julian Dates,
BJD$_{\rm TDB}$. To carry out this conversion, we used the tools made
available by \cite{Eastman2010}.

\section{White light curve: Data preparation and main analysis}
\label{Fitting}

\subsection{Construction of light curves and computation of spectro-photometric errors}
\label{WLC}

The two white light curves (OR1 and OR2) were constructed integrating
the fluxes of both target and reference stars in the wavelengths
between 520 and 750 nm (see shaded area on
Figure~\ref{fig:wav_vs_flux}). Limiting the wavelength region was
intended to minimize any impact related to the wavelength-dependency
of the pulsation amplitudes \citep[e.g.,][]{Breger2005}, and limb
darkening differences between the two data sets. Although some flux
was lost, this allowed us to carry out a coherent, simultaneous fit of
both transit and pulsations. The Fraunhofer and stellar lines
indicated with dashed vertical lines in Figure~\ref{fig:wav_vs_flux}
were used to check that the spectra were aligned within each night,
and among OR1 and OR2.

We computed the spectro-photometric errors using the formalism
provided by IRAF, as described in detail in Sect. 3.2 of
\cite{vonEssen2017}. In brief, uncertainties are computed using the
integrated fluxes within a given wavelength range and chosen aperture,
the respective area, the standard deviation of the sky region within
the same wavelength range, the number of sky pixels, and the gain of
the detector. Since errors determined in this way are known to be
slightly underestimated \citep{Gopal1995}, we scaled them up to meet
the standard deviation of the off-transit data points. In this way,
the spectro-photometric errors reflect not only the natural scatter of
the data, but also the expected increase of noise with increasing
airmass.

\cite{Huitson2017} reported the impact of wavelength-dependent
"stretches" in spectroscopic data over chromatic light curves, all
wrapped up in the context of transmission spectroscopy studies. The
stretches occur when a given pixel does not sample the same wavelength
in each exposure, mostly due to be mechanical or atmospheric
causes. The authors computed these stretches from cross-correlating
their GEMINI/GMOS spectra during transit solely around the Na and
H$\alpha$ lines. Their derived variability has an amplitude of about
one pixel (their Figure 2). As expected, the authors found this effect
to be more prominent for narrower wavelength bins than the ones they
were using ($\sim$10 nm). Even though they failed at reproducing this
variability by a physically-motivated model, they identify it to be
connected to the instrument response function. For response functions
steepening toward bluer wavelengths, the introduced flux changes
should be more prominent and, thus, increase the error on the derived
transmission spectrum.

To characterize this effect over our GTC/OSIRIS data and the potential
impact on our results, we carried out the same procedure as in
\cite{Huitson2017}. Our derived stretches perfectly mimic the
variability observed with FWHM, and the latter is found unimportant in
the data detrending process by the BIC (see
Section~\ref{sec:dmod}). Following the behaviour observed in
Figure~\ref{fig:environmental} (red points), the stretches for R500R
have no systematic noise. Only white noise around $\pm$1
\AA\ (equivalently, $\pm$0.2 pixels). In the case of R1000B (green
points in Figure~\ref{fig:environmental}) the stretches have a
systematic trend in agreement with the FWHM variability. However, the
amplitude is smaller as the spectra are of better quality and, thus,
the cross-correlation naturally improves. In this case, most of the
points lie around $\pm$0.3 \AA\ (equivalently, $\pm$ 0.1 pixels).  All
in all, our integration bands are double the size (20 vs. 10 nm),
which would dilute this effect by construction. Furthermore, our
derived stretches are on average an order of magnitude smaller. We did
not consider for our analysis of the transmission spectrum the edges
of the spectra, and in consequence we do not account in our analysis
fluxes showing the largest pixel-to-pixel variability. Thus, we
believe this effect is negligible. To avoid carrying out unnecessary
changes to an already complicated data set, in this work we do not
correct for the observed stretches.

From the derived light curves we first noticed a large scatter,
growing with increasing airmass. Due to short exposure times and the
large collecting area of the GTC, our photometric precision is
strongly limited by scintillation
\citep{Young1993,Kjeldsen1992}. However, owing to the large flux of
both stars the choice of the integration band does not significantly
affect the noise of the wavelength-binned light curves. Therefore, the
impact of choosing a narrower integration band to construct the white
light curves did not significantly diminished their overall
quality. To fit the transit light curves we take into account a model
with three main components: a transit, a detrending, and a pulsation
model.

\subsection{The transit model}

As primary transit model we use \citet{MandelAgol2002}'s
\texttt{occultquad}
routine\footnote{\url{http://www.astro.washington.edu/users/agol}}. The
parameters that we can infer are the semi-major axis in stellar radii,
$\mathrm{a/R_s}$, the orbital inclination, i, the orbital period, Per,
the mid-transit time, T$_0$, the planet-to-star radius ratio,
$\mathrm{R_P/R_S}$, and the limb darkening coefficients,
$\mathrm{u_1}$ and $\mathrm{u_2}$, from a quadratic limb-darkening
law.

Throughout this work, limb darkening coefficients are computed as
described in \cite{vonEssen2017} for specific wavelength regions, for
each transit, and for stellar parameters closely matching the ones of
\wasp. In order to properly compute the limb-darkening coefficients,
the efficiency of both CCD and grisms -along with their wavelength
dependency- have to be taken into consideration. Since the optical
setup differs between one transit and the other, simply because two
different grisms were used, we computed a set of linear and quadratic
limb darkening coefficients per observing run. For the white light
curve the derived values for OR1 are \mbox{u$_1$ = 0.333(1)} and
\mbox{u$_2$ = 0.245(6)}. For OR2, the limb darkening coefficients are
\mbox{u$_1$ = 0.322(9)} and \mbox{u$_2$ = 0.243(1)}. The similarity
between the two sets of LDs does not come as a surprise. While the
star and the integration wavelength region are exactly the same, the
responses of the two grisms are similar. For the derived values,
parenthesis indicate the precision in the fit of the stellar
intensities as a function of distance between the stellar limb and
center. As pointed out in \cite{vonEssen2017} and as observed by
\cite{Kervella2017}, the precision of the limb-darkening coefficients
only reflects a goodness of fit between models and a second order
polynomial, and not the true accuracy at which we know any
limb-darkening coefficients. The photometric precision of our data
does not allow to detect differences in limb darkening in the third
decimal. Therefore, for both white light curves we used as linear and
quadratic limb-darkening values their average. This means that we fit
only one transit model to the two data sets, simplifying the number of
parameters. Further considerations in the wavelength-dependent light
curves will be given in future sections. For \waspa, we considered
stellar atmospheric models provided by PHOENIX \citep{Peter1,Peter2}
for a surface gravity of 4.5, an effective temperature of \mbox{7400
  K}, and a metallicity of 0, closely matching \wasp's values reported
by \cite{CollierCameron2010} (\mbox{$\log(g) = 4.3 \pm 0.2$},
\mbox{$T_{eff}$ = 7430 $\pm$ 100 K}, and \mbox{[Fe/H] = 0.1 $\pm$
  0.2}, respectively).

Along this work, the limb darkening coefficients are considered as
fixed, and are computed in the same way described here per integration
band width. We fitted all the transit parameters, with the exception
of the orbital period \citep{vonEssen2014}, since its known to a high
degree of accuracy. For the transit parameters we considered Gaussian
priors using as starting values the ones derived in
\cite{vonEssen2014} (where the pulsations were accounted for) and as
width for their Gaussian priors five times their errors.

\subsection{Pulsation model}

As done in \cite{vonEssen2015}, to represent the stellar pulsations
our model comprises the sum of eight sinusoidal functions. Here, the
model parameters are the pulsation frequencies, amplitudes and
phases. To constrain the parameter space we used our prior knowledge
about the pulsation spectrum of the star, along with its behaviour in
time. For instance, in \cite{vonEssen2014} (see Section 3.5) we
detected a time-dependent change of the phases of the pulsations. In
order to characterize their shifts, we divided the data in monthly
groups and computed the shifts there. The closest two groups were
taken one month apart, which is approximately the time between OR1 and
OR2. Within this time lapse, some of the pulsations showed differences
in phases smaller than 0.2$\times$2$\pi$ (pulsations 1, 4, and 6),
while the others showed a phase shift with an amplitude up to
0.4$\times$2$\pi$. To both minimize the number of fitting parameters
and take the phase shifts into consideration, our pulsation model
comprises eight phases that are fitted simultaneously to the two
transits, plus eight phase shifts. We note, however, that both phases
and phase shifts can be degenerate. Thus, we fit them to the data
using Gaussian priors. Each shift has a Gaussian distribution around
zero, with standard deviations equal to the previously mentioned phase
shift amplitudes. Each phase has a Gaussian distribution around our
best initial guesses, with a standard deviation equal to 0.2. This
value is completely arbitrary. Since OR1 and OR2 are two years apart
from the data analyzed in \cite{vonEssen2014} and we don't know how
the phases continuously evolve in time, we don't know their current
values. Our initial guesses were thus computed evaluating the
pulsation model in different phase values, particularly dividing each
phase space in a grid ranging from 0 to 1 and a step of 0.05. Our
initial estimates for the phases are the ones that minimize the
standard deviation of the light curves once each pulsation model is
subtracted.

As the frequency resolution is 1/$\Delta$T \citep{Kurtz1983},
2$\times$4 hours of data are not sufficient to determine the
pulsations frequencies. Therefore, during our fitting procedure we
used the frequencies determined in \cite{vonEssen2014} as fixed
values. Furthermore, it is known that $\delta$ Scuti stars have
pulsation amplitudes that are also wavelength dependent
\citep{Breger1979,Breger2005}. The observations performed in
\cite{vonEssen2014} were taken mostly in the visible \mbox{($\sim$550
  $\pm$ 70 nm)} not matching with the wavelength coverage of the white
light curves. Since the overall amplitudes of the pulsations are small
and differential variability is not detectable from the photometric
precision of our data, at this stage, the amplitudes are considered as
fixed parameters with values equal to the ones derived in
\cite{vonEssen2014}. However, we account for any possible offset
between these and those from our observations by adding to our model
parameter space a constant amplitude offset per wavelength channel
equal to all the eight amplitudes, and to the two transits,
separately. In other words, while the transit parameters are fitted
simultaneously to the two light curves, the amplitudes in the
wavelength bins that match between OR1 and OR2 are completely
disconnected. However, if the wavelength range is the same for two
given light-curves between OR1 and OR2, as it is the case of the white
light curves, the resulting amplitude offsets should be consistent
between each other. The results obtained from this approach will serve
as consistency check of our methodology. All in all, the pulsation
model (PM) is as follows:

\begin{eqnarray}
    \mathrm{PM(t)} &=& \mathrm{PM(t)_{OR1} + PM(t)_{OR2}} \nonumber\\
                   &=& \sum_{i=1}^8 (A_i + A_j)\times \sin[2\pi(\nu_i t + \phi_i)] + \nonumber\\
                   & & \sum_{i=1}^8 (A_i + A_k)\times \sin[2\pi(\nu_i t + \phi_i + \Delta \phi_k)]\,.
\label{eq:eq1}
\end{eqnarray}

\noindent Here $\mathrm{A_i}$ and $\mathrm{\nu_i}$ correspond to the
eight amplitudes and frequencies found in \cite{vonEssen2014}
considered as fixed, $\mathrm{\phi_i}$ are the eight fitted phases,
and $\mathrm{A_j}$ and $\mathrm{A_k}$ are scaling factors that account
for wavelength-dependent amplitude differences. The indexes j and k
range between one and the number of wavelength bins of OR1 and OR2,
respectively. Thus, for the white light curves \mbox{j = 1} and
\mbox{k = 1}. $\mathrm{\Delta \phi_k}$ accounts for the phase shifts
between OR1 and OR2. The best-fit amplitudes and phases can be seen in
Table~\ref{tab:phases}. For both OR1 and OR2, the values of the phases
are absolute, and not differential. The amplitudes of OR1 and OR2 are
fully consistent within 1-$\sigma$ uncertainties, and the phase
differences are within the expected ranges. The errors of the phases
are purely statistical and are influenced by our choice of using
Gaussian priors. Thus, they should be considered with care.

\subsection{Detrending model}
\label{sec:dmod}

To account for systematic noise in our data, we tested several
dependencies. Due to the deformations caused by the pulsations it is
hard to visually inspect the success of our detrending strategy. Thus,
to choose the amount of detrending components we made use of the
minimization of the Bayesian Information Criterion, \mbox{BIC =
  $\chi^2$ + k ln N}. For the BIC, k is the number of model parameters
and N is the number of data points. Besides the linear and quadratic
time-dependent polynomials, we used a linear combination of airmass
(also considered quadratically), FWHM, background counts and spatial
position, combining them in several ways. The detrending function
minimizing the BIC corresponds to a quadratic function of airmass plus
a linear function of background counts. This coincides with the
variability observed in Figure~\ref{fig:environmental}. The FWHM and
the spatial position, not favored by the BIC, are actually the ones
that vary the less along the ORs. The detrending model (DM) looks as
follows:

\begin{equation}
    \mathrm{DM(t)} = c_0 + c_1 \times \chi + c_2    \times \chi^2 + c_3 \times bc,\,
\end{equation} 

\noindent where $c_0$, $c_1$, $c_2,$ and $c_3$ are the detrending
coefficients, $\chi$ corresponds to the airmass, and bc to the
background counts. The four detrending coefficients have all uniform
priors. Since stellar light suffers a wavelength-dependent absorption
when crossing our atmosphere, a set of four detrending coefficients is
fitted to each chromatic light curve individually.

\subsection{Computation of model parameters}

\begin{figure}[ht!]
  \centering
  \includegraphics[width=.5\textwidth]{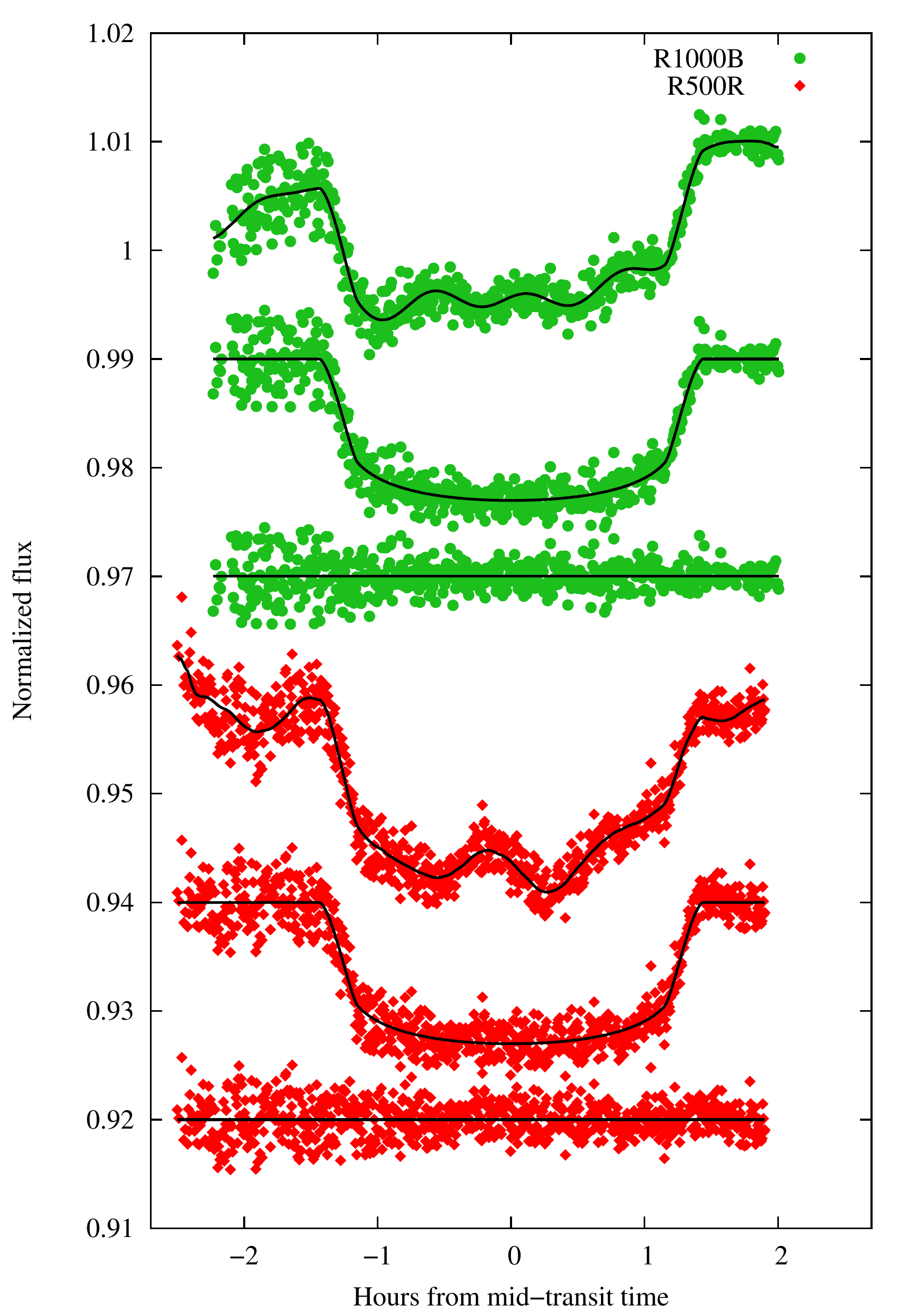}
  \caption{\label{fig:WLC} White light curves of \waspb\ obtained with
    the GTC, artificially shifted for visual inspection. Green filled
    circles correspond to OR1, and red filled diamonds to OR2. Each OR
    is presented in three sections, described from top to bottom as
    follows: the raw photometry along with the best-fit model in black
    continuous line, the transit light curve after the pulsations and
    detrending models are subtracted, along with the best-fit transit
    model in black continuous line, and the residual light curves on
    the bottom.}
\end{figure}

To derive the model parameters we fitted the two white light curves
simultaneously using a Markov-chain Monte Carlo (MCMC) approach, all
wrapped up in
PyAstronomy\footnote{www.hs.uni-hamburg.de/DE/Ins/Per/Czesla/PyA/PyA/index.html},
a collection of Python routines providing a convenient interface for
fitting and sampling algorithms implemented in the PyMC
\citep{Patil2010} and SciPy \citep{Jones2001} packages. After
\mbox{1$\times$10$^6$} iterations and a burn-in of the initial
\mbox{2$\times$10$^5$} samples, we computed the mean and standard
deviation \mbox{(1-$\sigma$)} of the posterior distributions of the
parameters and used them as best-fit values and uncertainties,
respectively. The best-fit solutions, along with their errors, are
displayed in Table~\ref{tab:WLC}. To check for convergence of the MCMC
chains we divided them in three equally large subgroups and computed
their respective mean and standard deviations. We considered a chain
to converge if these values were consistent between each other at
1-$\sigma$ level. We finished up with a visual inspection of the
chains. The transit photometry corresponding to OR1 and OR2, the
best-fit model, the detrended light curves and the residual light
curves can be seen in Figure~\ref{fig:WLC}. The transit parameters
derived in this work show smaller uncertainties than bibliographic
values.

\begin{table*}[ht!]
  \centering
  \caption{\label{tab:WLC}Best-fit parameters derived from the two
    white light curves, compared to the values reported in \cite{vonEssen2014}.}
  \begin{tabular}{l c c}
    \hline \hline
    Parameter           &        OR1/OR2           &   \cite{vonEssen2014} \\
    \hline
    a/R$_s$             & 3.62 $\pm$ 0.02          &  3.68 $\pm$ 0.03   \\ 
    i ($^{\circ}$)      & 88.24 $\pm$ 0.28         &  87.90 $\pm$ 0.93    \\
    R$_P$/R$_s$         & 0.1053 $\pm$ 0.0004      &  0.1046 $\pm$ 0.0006  \\
    T$_0$ (BJD$_{TDB}$) & 1878.65739 $\pm$ 0.00015 &  507.5222 $\pm$ 0.0003 \\
    Per (days)          & 1.2198675 (fixed)        & 1.2198675 $\pm$1.1$\times$10$^{-6}$ \\
    u$_1$               & 0.327(8) (fixed)         &  (see Table 8, \cite{vonEssen2014})    \\
    u$_2$               & 0.244(3) (fixed)         &  (see Table 8, \cite{vonEssen2014})    \\
    \hline
  \end{tabular}   
  \tablefoot{T$_0$ is given in BJD$_{TDB}$ - 2450000.}
\end{table*}

\begin{table}[ht!]
  \centering
  \caption{\label{tab:phases} Best-fit phases, $\phi_i$, and amplitudes, $\mathrm{A_j}$ and $\mathrm{A_k}$, derived from the two white light curves. Amplitudes are given in parts-per-thousand (ppt), phases in units of 2$\pi$, and the subindices follow the notation of Eq.~\ref{eq:eq1}. }
  \begin{tabular}{l c c}
    \hline \hline
    Parameter           &        OR1           &   OR2 \\
    \hline
    $\phi_{\mathrm{1}}$      & 0.334  $\pm$ 0.005   &   0.676  $\pm$ 0.005   \\
    $\phi_{\mathrm{2}}$      & 0.644  $\pm$ 0.005   &   0.334  $\pm$ 0.005   \\
    $\phi_{\mathrm{3}}$      & 0.867  $\pm$ 0.005   &   0.583  $\pm$ 0.005   \\
    $\phi_{\mathrm{4}}$      & 0.114  $\pm$ 0.005   &   0.984  $\pm$ 0.005   \\
    $\phi_{\mathrm{5}}$      & 0.843  $\pm$ 0.005   &   0.462  $\pm$ 0.005   \\
    $\phi_{\mathrm{6}}$      & 0.089  $\pm$ 0.005   &   0.116  $\pm$ 0.005   \\
    $\phi_{\mathrm{7}}$      & 0.323  $\pm$ 0.005   &   0.475  $\pm$ 0.005   \\
    $\phi_{\mathrm{8}}$      & 0.356  $\pm$ 0.005   &   0.691  $\pm$ 0.005   \\
    A$_{j/k}$                &  0.22 $\pm$ 0.05     &   0.19 $\pm$ 0.05  \\
    \hline
  \end{tabular}
\end{table}

\subsection{Treatment of correlated noise}

Although our model accounts for the deformations produced by the
pulsations, these are actually the ones that do not allow us to
visually assess whether the detrending model is sufficient to describe
our data. At this stage, we reply purely on statistical tools such as
the BIC. To account for as much as possible either systematic effects
not accounted for in our model, or a poor pulsation model (i.e., not
enough pulsation frequencies yet discovered), we carried out the MCMC
fitting process twice. After the first MCMC run that was carried out
as described in the previous section, we computed residual light
curves for OR1 and OR2 subtracting our best-fit models to each
photometric light curve. Following the approach described in
\cite{Gillon2006,Winn2008,Carter2009}, we computed the $\beta$-value
from the residual light curves following their prescription, averaging
$\beta$'s calculated within time bins equal to 0.8, 0.9, 1, 1.1, and
1.2 times the transit ingress/egress time. Then, we enlarged the
spectro-photometric error bars by the $\beta$'s, and we re-run MCMC to
re-compute the best-fit parameters. The derived values are reported in
Table~\ref{tab:WLC}, and the $\beta$ values are \mbox{$\beta_{OR1}$ =
  1.05}, and \mbox{$\beta_{OR2}$ = 1.29}.

\subsection{Construction of the wavelength-binned light curves and corresponding fitting parameters}

\begin{figure}[ht!]
  \centering
  \includegraphics[width=.5\textwidth]{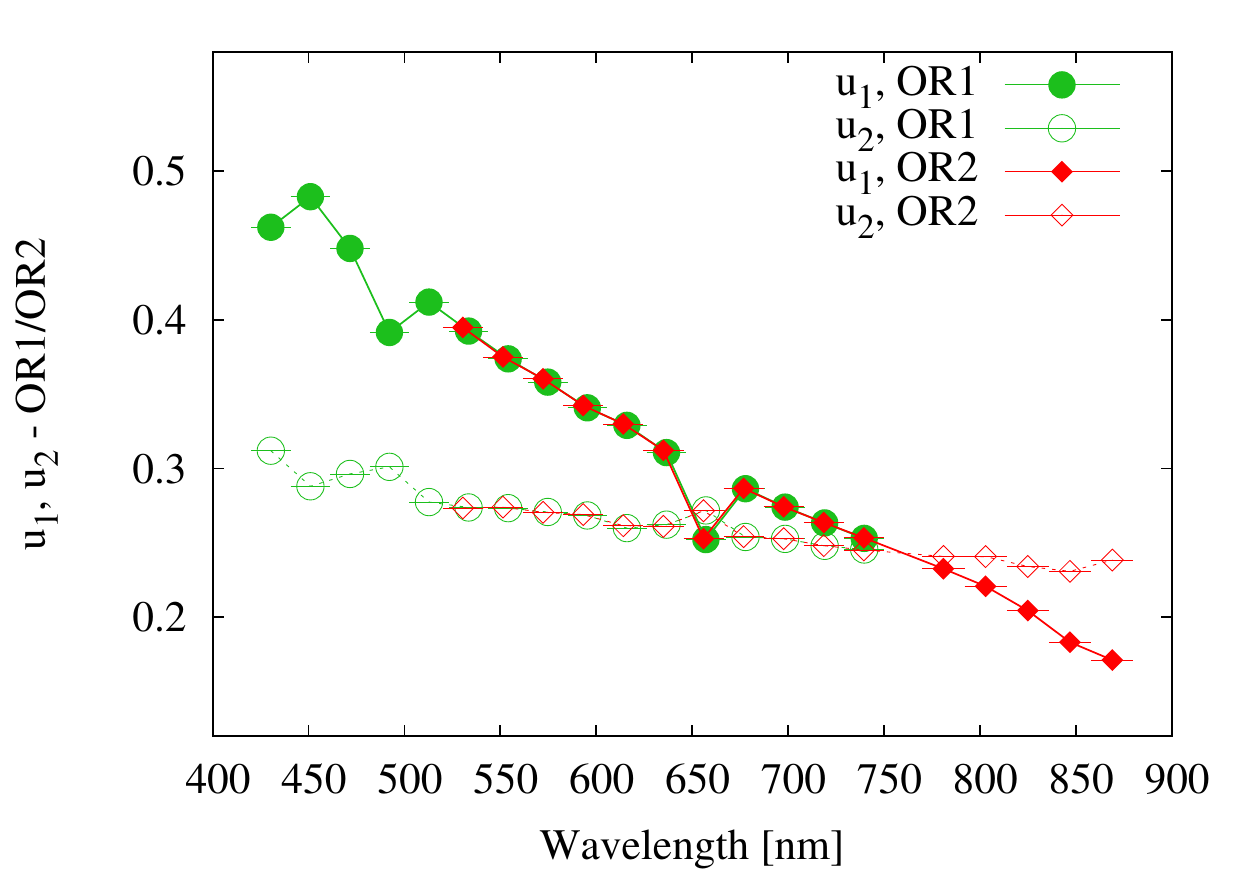}
  \caption{\label{fig:LDs} Limb-darkening coefficients, u$_1$ and
    u$_2$, for OR1 in green circles and OR2 in red diamonds. The
    linear and quadratic coefficients are plotted in filled and empty
    points, respectively.}
\end{figure}

To derive the transmission spectrum of \waspb\ from color-dependent
light curves, we carried out a similar approach than the one of the
white light curve regarding the computation of spectro-photometric
errors, their respective scaling to meet the standard deviation, the
computation of correlated noise, and the calculation of limb-darkening
coefficients. To obtain the best-fit planet-to-star radius ratio as a
function of spectral bin we made use of MCMC in the exact same fashion
as with the white light curves. The values of planet-to-star radii
ratio reported in this work are thus obtained from the second MCMC run
of \mbox{1$\times$10$^6$} iterations, after a burn-in phase of the
first \mbox{2$\times$10$^5$} samples. To compute the chromatic light
curves we divided the total spectral coverage in wavelength bins with
a width of $\sim$22 nm. Figure~\ref{fig:LDs} shows the derived linear
(full green circles) and quadratic (empty green circles)
limb-darkening coefficients for OR1, along with the corresponding ones
for OR2 (filled and empty red diamonds, respectively). As the figure
reveals, there is an almost identical match between the limb-darkening
values derived from the two observing runs in the wavelength regions
where they coincide. The small differences are caused by the slightly
different responses of the used grisms, R1000B and R500R.

The strength at which absorption of stellar light takes place in our
atmosphere strongly depends on wavelength and airmass. For both OR1
and OR2, airmass values at the beginning of the observations were
slightly larger than 2. Therefore, our chromatic light curves have a
differential deformation that can clearly be distinguished by visually
inspecting the raw light curves. To account for this, we fit the four
detrending coefficients to each light curve independently. We also
tested a common mode correction by dividing the chromatic light curves
by the white light curve residuals \citep[see
  e.g.,][]{Lendl2016,Nikolov2016,Gibson2017}. However, this exercise
did not reduce the noise in the chromatic light curves nor did it
notably change the results. Therefore, it was finally not included in
the analysis procedure.

In order to properly propagate the errors on the transit parameters to
the chromatic light curves, rather than fixing the transit parameters
derived from the white light curves these were set free. In all cases
the planet-to-star radius ratio, R$_\mathrm{P}$/R$_\mathrm{S}$, was
fitted considering a uniform probability density function limited
between 0 and 0.3, way above and below bibliographic values. The
semi-major axis, inclination and mid-transit time were fitted to the
chromatic light curves using Gaussian probability density functions
with mean and standard deviation equal to the best-fit and error
values obtained from the white light curves, respectively. These
parameters do not depend on wavelength. In consequence, we treated
them as equal during the MCMC fitting.

\begin{figure}[ht!]
  \centering
  \includegraphics[width=.5\textwidth]{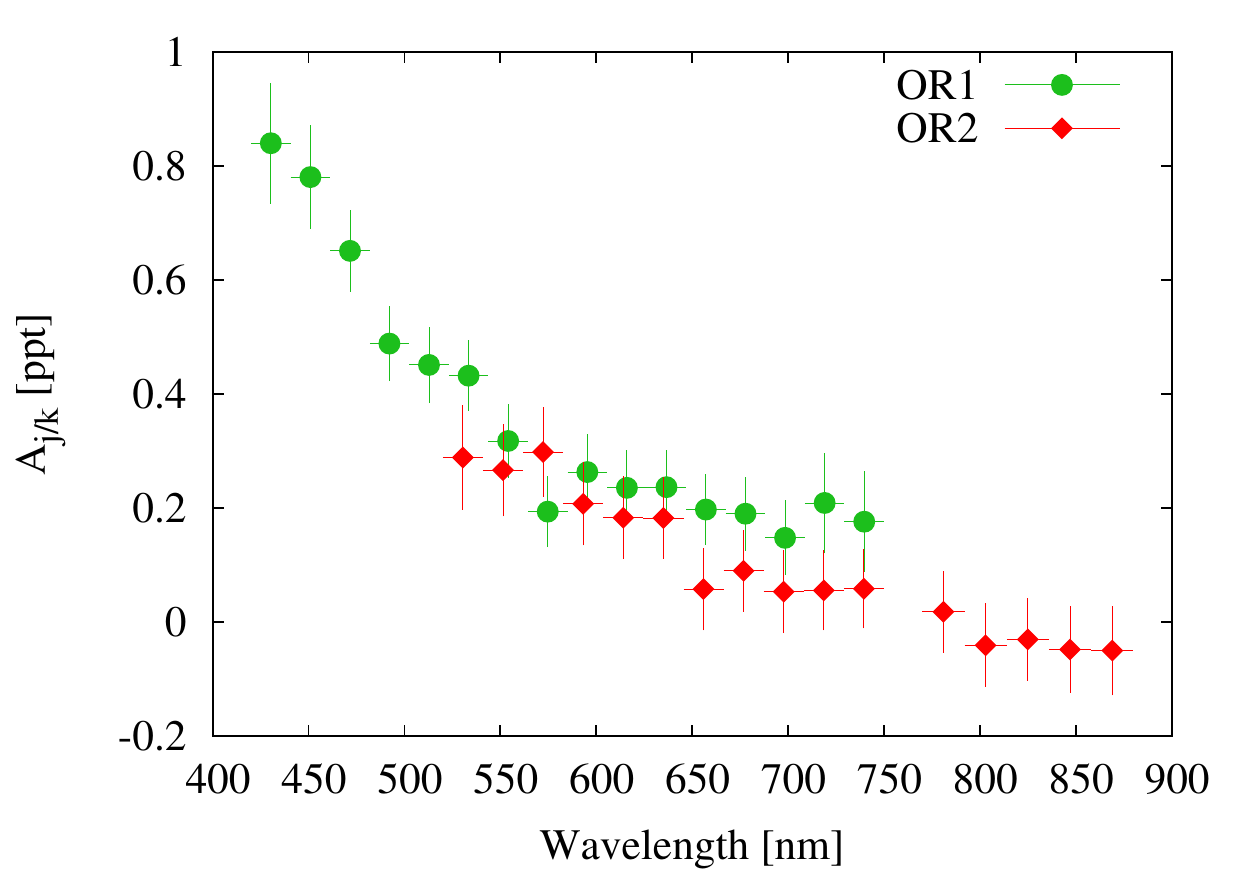}
  \caption{\label{fig:amplitudes} Amplitude offset, A$_{j/k}$, as a function of wavelength channel. Here, j represents the wavelength channels of OR1, and k the ones of OR2. Best-fit values and 1-$\sigma$ uncertainties are plotted in green filled circles for OR1, and in red filled diamonds for OR2.}
\end{figure}

Equivalently to the transit parameters, rather than fixing the sixteen phases to the values obtained from the white light curves, these were fitted to the chromatic light curves simultaneously. Thus, rather than having (8$\times$2)$\times$n fitting phases, where n is the number of wavelength channels, we fitted only 8x2. As initial values we used the ones obtained from the white light curve analysis, along with a Gaussian probability density function with standard deviation equal to the reported 1-$\sigma$ errors. Finally, to account for wavelength-dependent variations of the pulsation amplitudes we considered the previously mentioned amplitude offset, A$_{j/k}$. We fitted one amplitude offset to each chromatic light curve, using uniform priors between -0.1 and 0.1 ppt. Pulsation amplitudes are expected to grow with stellar emission. Since \wasp\ emits most of its flux in the bluer wavelengths, it is expected for A to grow with decreasing wavelength. Figure~\ref{fig:amplitudes} shows the best-fit values of A$_{j/k}$, along with 1-$\sigma$ uncertainties for OR1 (green filled circles) and OR2 (red filled diamonds). The figure shows two very important results. On one hand, as expected in $\delta$ Scuti stars the amplitude offset relative to the amplitudes found in \cite{vonEssen2014} does decrease with increasing wavelength. On the other hand, even though the pulsation amplitudes were fitted to each transit with uniform priors individually, within 1-$\sigma$ errors they perfectly overlap in the common wavelength region.

\begin{figure*}
  \centering
  \includegraphics[width=.9\textwidth]{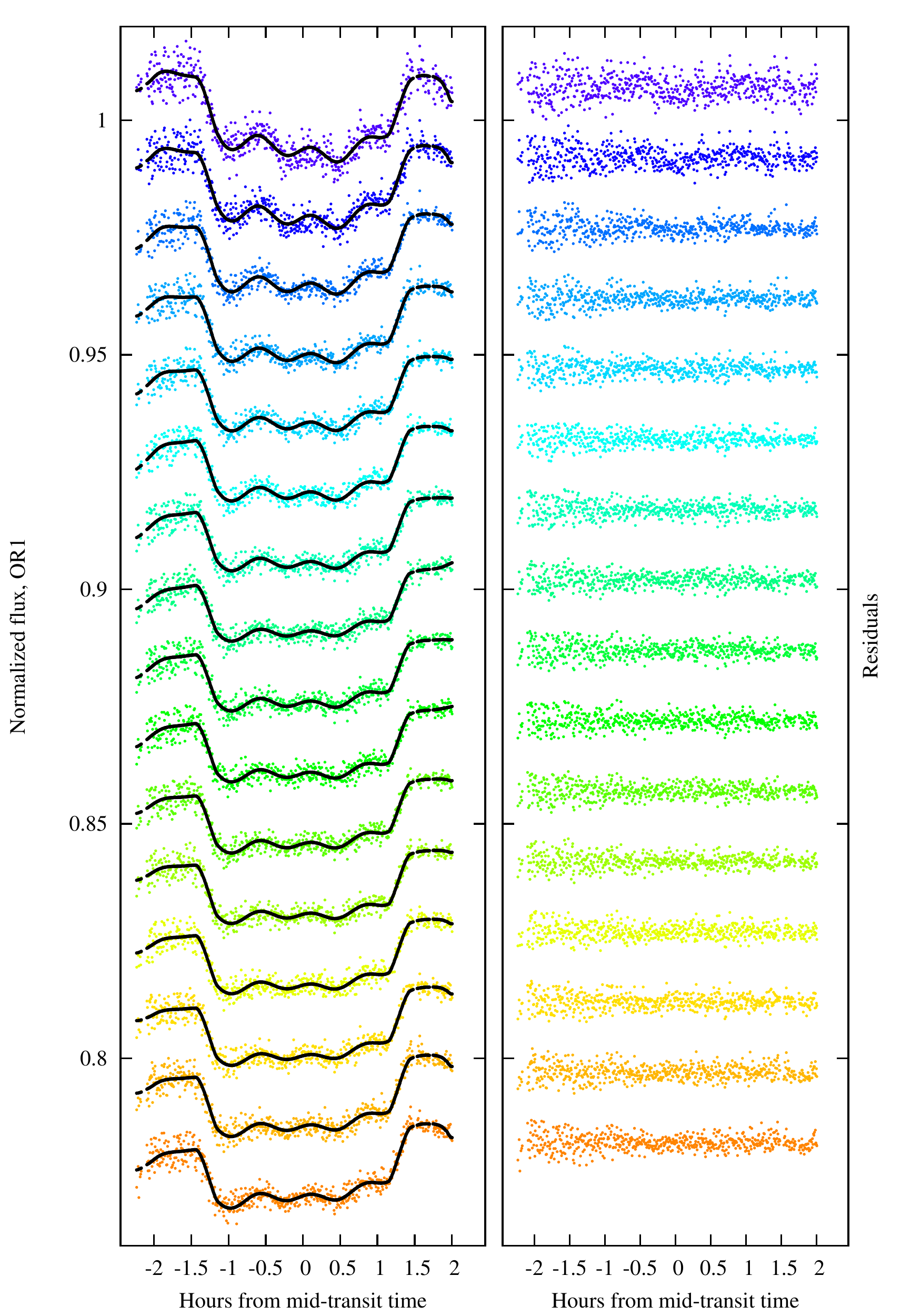}
  \caption{\label{fig:OR1_LCs} Our 16 chromatic transit light curves
    obtained during OR1 on the left panels, and corresponding
    residuals on the right. The continuous black line represents our
    model, comprising the pulsations, the detrending, and the transit
    models. Transits are plotted in hours from mid-transit time, and
    are shifted vertically to allow for visual inspection.}
\end{figure*}

\begin{figure*}
  \centering
  \includegraphics[width=.9\textwidth]{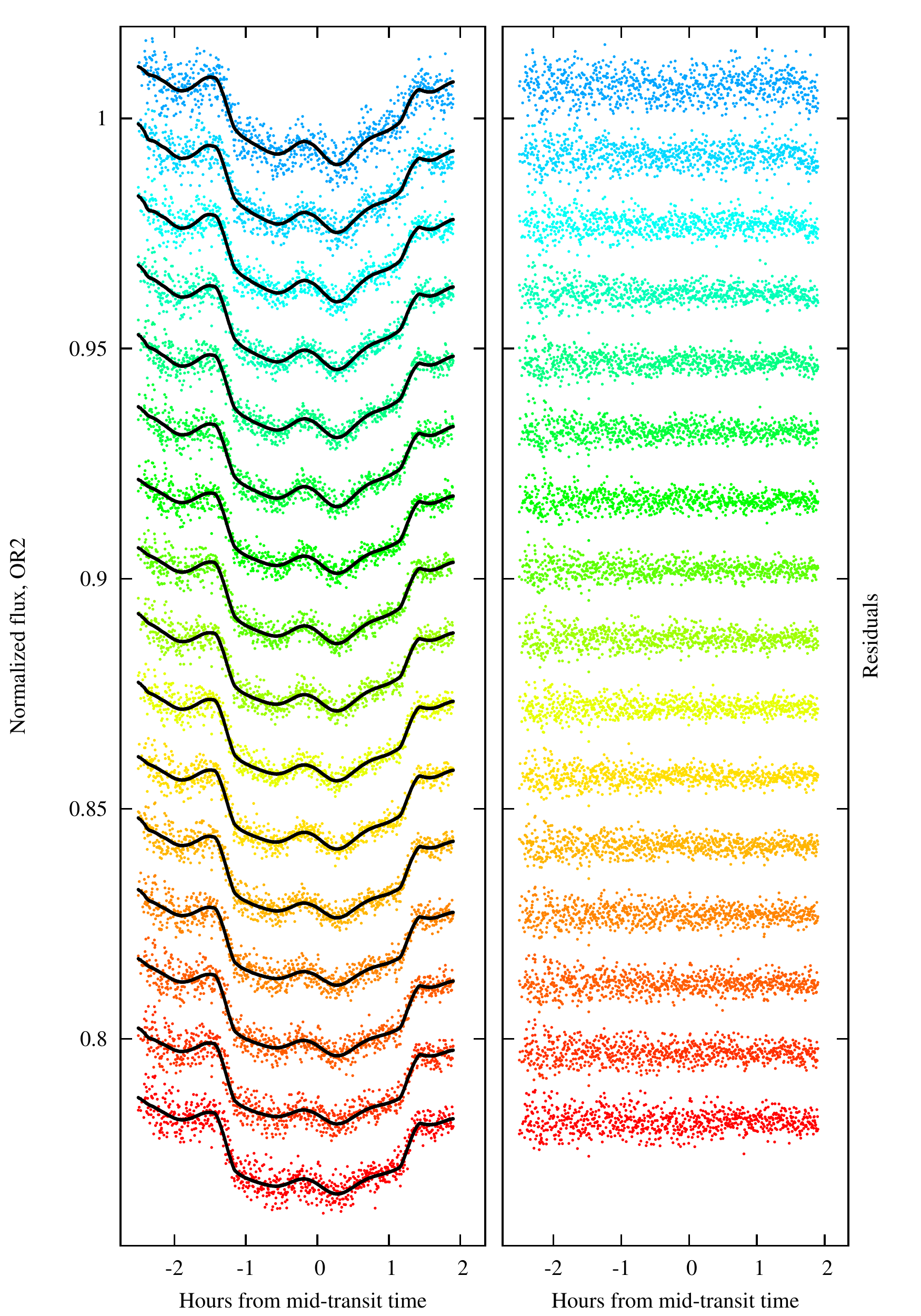}
  \caption{\label{fig:OR2_LCs} Results obtained during OR2 (see
    Figure~\ref{fig:OR1_LCs}).}
\end{figure*}

Figure~\ref{fig:OR1_LCs} and Figure~\ref{fig:OR2_LCs} show the
chromatic light curves of \waspb, shifted with respect to the best-fit
mid-transit time and shown in hours. The light curves are color-coded
with respect to the wavelength channel. Black continuous line shows
the best-fit model, which comprises the transit, the detrending, and
the pulsation models. Photometric errors are enlarged by their
respective $\beta$ values. The chromatic light curves have been
vertically shifted to allow for visual
inspection. Table~\ref{tab:results} specifies the wavelength channels,
the derived planet-to-star radius along with their uncertainties, the
limb darkening values, and the best-fit detrending coefficients.

\begin{table*}
  \centering
  \caption{\label{tab:results} From left to right, center of
    integration band and integration width, best fit planet-to-star
    radius ratio, detrending coefficients $\mathrm{c_0}$,
    $\mathrm{c_1}$, $\mathrm{c_2}$, and $\mathrm{c_3}$, standard
    deviation of residuals in ppt, and $\beta$ values. Best-fit values
    are given with 1-$\sigma$ uncertainties.}
  \begin{tabular}{l l l l l l l l}
    \hline \hline
    $\lambda_o \pm \Delta \lambda$ (nm)  &    $\mathrm{R_P/R_s}$  &  $\mathrm{c_0}$   &   $\mathrm{c_1}$  &  $\mathrm{c_2}$  &  $\mathrm{c_3}$   & $\sigma$ (ppt) & $\beta$  \\
    \hline
    OR1 & & & & & \\
    \hline
    430.3 $\pm$ 10.3 & 0.1082 $\pm$ 0.0014 & 1.0091 $\pm$ 0.0003 & 0.025 $\pm$ 0.003 &-0.020 $\pm$ 0.004  & -0.67 $\pm$ 0.09 & 2.4 & 1.63 \\
    450.9 $\pm$ 10.3 & 0.1070 $\pm$ 0.0012 & 1.0085 $\pm$ 0.0003 & 0.015 $\pm$ 0.002 &-0.012 $\pm$ 0.003  & -0.50 $\pm$ 0.07 & 2.0 & 1.69 \\
    471.6 $\pm$ 10.3 & 0.1073 $\pm$ 0.0009 & 1.0085 $\pm$ 0.0002 & 0.007 $\pm$ 0.002 &-0.009 $\pm$ 0.003  & -0.35 $\pm$ 0.06 & 1.6 & 1.66 \\
    492.2 $\pm$ 10.3 & 0.1080 $\pm$ 0.0009 & 1.0083 $\pm$ 0.0002 & 0.003 $\pm$ 0.002 &-0.006 $\pm$ 0.002  & -0.24 $\pm$ 0.05 & 1.5 & 1.15 \\
    512.8 $\pm$ 10.3 & 0.1067 $\pm$ 0.0009 & 1.0082 $\pm$ 0.0002 & 0.000 $\pm$ 0.002 &-0.007 $\pm$ 0.002  & -0.17 $\pm$ 0.05 & 1.5 & 1.37 \\
    533.4 $\pm$ 10.3 & 0.1067 $\pm$ 0.0008 & 1.0084 $\pm$ 0.0002 & 0.000 $\pm$ 0.002 &-0.011 $\pm$ 0.002  & -0.21 $\pm$ 0.05 & 1.4 & 1.28 \\
    554.1 $\pm$ 10.3 & 0.1057 $\pm$ 0.0009 & 1.0080 $\pm$ 0.0002 & -0.004 $\pm$ 0.002&-0.005 $\pm$ 0.002  & -0.10 $\pm$ 0.05 & 1.5 & 1.01 \\
    574.7 $\pm$ 10.3 & 0.1048 $\pm$ 0.0009 & 1.0075 $\pm$ 0.0002 & -0.010 $\pm$ 0.002& 0.001 $\pm$ 0.002  &  0.04 $\pm$ 0.05 & 1.4 & 1.20 \\
    595.3 $\pm$ 10.3 & 0.1037 $\pm$ 0.0009 & 1.0076 $\pm$ 0.0002 & -0.004 $\pm$ 0.002&-0.003 $\pm$ 0.002  & -0.09 $\pm$ 0.05 & 1.5 & 1.05 \\
    615.9 $\pm$ 10.3 & 0.1060 $\pm$ 0.0009 & 1.0078 $\pm$ 0.0002 & -0.006 $\pm$ 0.002&-0.002 $\pm$ 0.002  & -0.02 $\pm$ 0.05 & 1.5 & 1.06 \\
    636.6 $\pm$ 10.3 & 0.1041 $\pm$ 0.0009 & 1.0074 $\pm$ 0.0002 & -0.004 $\pm$ 0.002& 0.001 $\pm$ 0.002  & -0.13 $\pm$ 0.05 & 1.5 & 1.07 \\
    657.2 $\pm$ 10.3 & 0.1048 $\pm$ 0.0009 & 1.0073 $\pm$ 0.0002 & -0.003 $\pm$ 0.002& 0.002 $\pm$ 0.002  & -0.13 $\pm$ 0.05 & 1.4 & 1.07 \\
    677.8 $\pm$ 10.3 & 0.1049 $\pm$ 0.0009 & 1.0075 $\pm$ 0.0002 & -0.002 $\pm$ 0.002&-0.001 $\pm$ 0.002  & -0.18 $\pm$ 0.05 & 1.4 & 1.08 \\
    698.4 $\pm$ 10.3 & 0.1043 $\pm$ 0.0009 & 1.0072 $\pm$ 0.0002 & -0.002 $\pm$ 0.002& 0.004 $\pm$ 0.002  & -0.23 $\pm$ 0.05 & 1.5 & 1.12 \\
    719.1 $\pm$ 10.3 & 0.1057 $\pm$ 0.0009 & 1.0076 $\pm$ 0.0002 & 0.001 $\pm$ 0.002 &-0.001 $\pm$ 0.002  & -0.33 $\pm$ 0.05 & 1.5 & 1.11 \\
    739.7 $\pm$ 10.3 & 0.1055 $\pm$ 0.0009 & 1.0076 $\pm$ 0.0002 & 0.002 $\pm$ 0.002 &-0.004 $\pm$ 0.002  & -0.39 $\pm$ 0.05 & 1.6 & 1.06 \\
    \hline
    OR2 & & & & & \\
    \hline
    530.5 $\pm$ 10.5 & 0.1046 $\pm$ 0.0021 & 1.0061 $\pm$ 0.0003 & 0.010 $\pm$ 0.005 & 0.010 $\pm$ 0.003  & -0.58 $\pm$ 0.65 & 3.0 & 1.16 \\
    551.4 $\pm$ 10.5 & 0.1037 $\pm$ 0.0013 & 1.0060 $\pm$ 0.0003 & 0.016 $\pm$ 0.004 & 0.016 $\pm$ 0.003  & -1.31 $\pm$ 0.54 & 2.4 & 1.03 \\
    572.3 $\pm$ 10.5 & 0.1037 $\pm$ 0.0013 & 1.0061 $\pm$ 0.0003 & 0.014 $\pm$ 0.004 & 0.014 $\pm$ 0.003  & -1.20 $\pm$ 0.56 & 2.1 & 1.05 \\
    593.2 $\pm$ 10.5 & 0.1044 $\pm$ 0.0012 & 1.0060 $\pm$ 0.0002 & 0.019 $\pm$ 0.004 & 0.019 $\pm$ 0.003  & -0.74 $\pm$ 0.51 & 1.9 & 1.06 \\
    614.1 $\pm$ 10.5 & 0.1043 $\pm$ 0.0012 & 1.0060 $\pm$ 0.0002 & 0.018 $\pm$ 0.004 & 0.018 $\pm$ 0.003  & -0.79 $\pm$ 0.57 & 1.8 & 1.04 \\
    635.0 $\pm$ 10.5 & 0.1033 $\pm$ 0.0012 & 1.0059 $\pm$ 0.0002 & 0.016 $\pm$ 0.004 & 0.016 $\pm$ 0.003  & -0.43 $\pm$ 0.48 & 1.8 & 1.26 \\
    655.9 $\pm$ 10.5 & 0.1049 $\pm$ 0.0012 & 1.0063 $\pm$ 0.0002 & 0.013 $\pm$ 0.004 & 0.013 $\pm$ 0.003  & -0.27 $\pm$ 0.52 & 1.8 & 1.05 \\
    676.8 $\pm$ 10.5 & 0.1063 $\pm$ 0.0011 & 1.0064 $\pm$ 0.0002 & 0.016 $\pm$ 0.004 & 0.016 $\pm$ 0.002  & -0.26 $\pm$ 0.52 & 1.7 & 1.05 \\
    697.7 $\pm$ 10.5 & 0.1044 $\pm$ 0.0012 & 1.0060 $\pm$ 0.0002 & 0.019 $\pm$ 0.004 & 0.019 $\pm$ 0.003  & -0.36 $\pm$ 0.50 & 1.8 & 1.01 \\
    718.6 $\pm$ 10.5 & 0.1072 $\pm$ 0.0011 & 1.0064 $\pm$ 0.0002 & 0.016 $\pm$ 0.004 & 0.016 $\pm$ 0.003  & -0.59 $\pm$ 0.50 & 1.7 & 1.08 \\
    739.5 $\pm$ 10.5 & 0.1058 $\pm$ 0.0011 & 1.0062 $\pm$ 0.0002 & 0.015 $\pm$ 0.004 & 0.015 $\pm$ 0.003  & -0.28 $\pm$ 0.52 & 1.7 & 1.07 \\
    781.0 $\pm$ 11.0 & 0.1073 $\pm$ 0.0012 & 1.0065 $\pm$ 0.0002 & 0.014 $\pm$ 0.004 & 0.014 $\pm$ 0.003  & -0.79 $\pm$ 0.52 & 1.7 & 1.09 \\
    803.0 $\pm$ 11.0 & 0.1054 $\pm$ 0.0012 & 1.0062 $\pm$ 0.0002 & 0.013 $\pm$ 0.004 & 0.013 $\pm$ 0.003  & -0.64 $\pm$ 0.54 & 1.9 & 1.03 \\
    825.0 $\pm$ 11.0 & 0.1069 $\pm$ 0.0012 & 1.0065 $\pm$ 0.0002 & 0.011 $\pm$ 0.004 & 0.011 $\pm$ 0.003  & -0.25 $\pm$ 0.49 & 1.9 & 1.05 \\
    847.0 $\pm$ 11.0 & 0.1078 $\pm$ 0.0013 & 1.0067 $\pm$ 0.0003 & 0.010 $\pm$ 0.004 & 0.010 $\pm$ 0.003  & -0.61 $\pm$ 0.50 & 2.0 & 1.04 \\
    869.0 $\pm$ 11.0 & 0.1082 $\pm$ 0.0013 & 1.0066 $\pm$ 0.0003 & 0.011 $\pm$ 0.004 & 0.011 $\pm$ 0.003  & -0.19 $\pm$ 0.58 & 2.3 & 1.39 \\
    \hline
  \end{tabular}
\end{table*}

\subsection{Consistency checkup of our results}

We identify three critical components of our model and model strategy
that might have an effect on the derived transmission spectrum of
\waspb. These are the selection of particular detrending components,
the specific integration bandwidths, and the particular way we treat
the pulsations.

To investigate the impact of the choice of detrending model into the
determination of the transmission spectrum of \waspb, we carried out
the same procedure but as for the detrending models, we considered the
following:

\begin{enumerate}
    \item A first order, time-dependent polynomial,\\ DM(t) = $c_0$ + $c_1$t.
    \item A second order, time-dependent polynomial,\\ DM(t) = $c_0$ + $c_1$t + $c_2$t$^2$.
    \item A linear relation with airmass and background,\\ DM($\chi$, bc) = $c_0$ + $c_1\chi$ + $c_2$bg.
    \item A linear relation with airmass, and quadratic with background,\\ DM($\chi$, bc) = $c_0$ + $c_1\chi$ + $c_2$bc + $c_3$bc$^2$.
    \item A quadratic relation with airmass,\\ DM($\chi$) = $c_0$ + $c_1\chi$ + $c_2\chi^2$.
\end{enumerate}

\noindent To investigate if the choice of integration bandwidth has an
impact in the transmission spectrum of \waspb, we changed their width
and re-computed new light curves. For each detrending function and
choice of bandwidth, we fitted the chromatic light curves in the exact
same fashion as described in previous sections.

\begin{figure*}[ht!]
  \centering
  \includegraphics[width=.45\textwidth, angle=270]{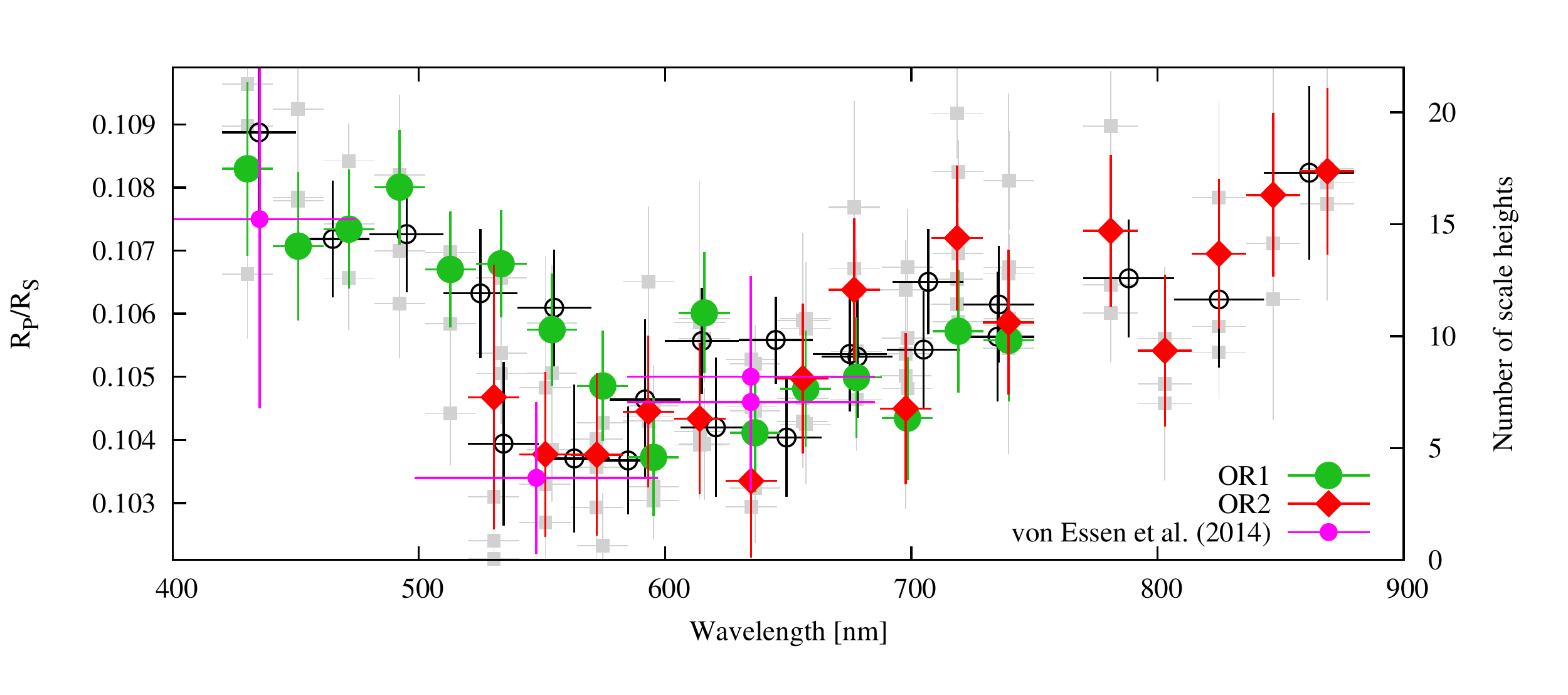}
  \caption{\label{fig:RpRs} Transmission spectrum of \waspb. Green
    circles and red diamonds correspond to the planet-to-star radius
    ratio obtained from OR1 and OR2, respectively. Gray light squares
    are derived fitting different detrending components to the
    data. Pink filled circles are from \cite{vonEssen2014}, and open
    black circles are considering larger integration bins. In all
    cases, error bars are at 1-$\sigma$ level and horizontal lines
    indicate the size of the wavelength bin.}
\end{figure*}

Our results are shown in Figure~\ref{fig:RpRs}. The green circles and
red diamonds correspond to the OR1 and OR2, respectively. Over-plotted
as gray squares and black empty circles we show the results derived
from different detrending components and bandwidths,
respectively. Pink filled circles show values reported in
\cite{vonEssen2014}. As the figure shows, our results are fully
consistent at 1-$\sigma$ uncertainties.  In addition, similarly to
\cite{vonEssen2015} we also considered additional models for the
stellar pulsations, included in our data analysis considering the
following changes:

\begin{enumerate}
\item Eight frequencies fitted with Gaussian priors, using our best-fit values from \cite{vonEssen2014} and their respective errors as mean and standard deviations for the priors.
\item A unique set of eight phases for the two transits, fitted to the data with uniform priors.
\item A unique set of planet-to-star radius ratio for each wavelength bin where the two transits coincide.
\item A unique set of amplitude offsets for each wavelength bin where the two transits coincide.
\item Two sets of phases with uniform priors.
\end{enumerate}

In this analysis, point 1. returned consistent results for
$\mathrm{R_P/R_S}$ but with slightly larger error bars. So did points
2. and 3., but with slightly smaller error bars where the wavelengths
coincided. A visual inspection of the residuals obtained from point
4. repeated systematically using different detrending functions showed
clear signs of pulsations not properly accounted for. For example,
when fitting the two white light curves with two set of phases, as
stated in previous sections the derived beta-values were
\mbox{$\beta_{OR1}$ = 1.05}, and \mbox{$\beta_{OR2}$ = 1.29}. After
considering one set of phases and recomputing the beta-values, we
obtained \mbox{$\beta_{OR1,min}$ = 1.74}, \mbox{$\beta_{OR1,max}$ =
  6.93}, \mbox{$\beta_{OR2,min}$ = 2.16}, and \mbox{$\beta_{OR2,max}$
  = 7.03} as minimum and maximum values of our best-fit models. The
corresponding BIC values were in all cases larger than 1700. This
supports the need for the two sets of phases.

\section{Transmission spectrum of \waspb}
\label{TS}

\subsection{Atmospheric retrieval}
\label{sec_retriev}

We performed an atmospheric retrieval on the combined spectrum of
WASP-33b to constrain the atmospheric properties at the day-night
terminator region. We employed an atmospheric retrieval code for
transit spectroscopy adapted from the recent work of
\citet{2018MNRAS.474..271G}. Our code computes line-by-line radiative
transfer in a transmission geometry and assumes a plane parallel
planetary atmosphere in hydrostatic equilibrium. The model
parametrizes the pressure-temperature profile of the atmosphere using
the prescription of \citet{2009ApJ...707...24M} containing six free
parameters. The volumetric mixing ratios of the chemical species in
the atmosphere are also free parameters in the retrieval framework. We
employed chemical opacities adopted from the work of
\citet{2017MNRAS.472.2334G, 2018MNRAS.474..271G}. Our models consider
molecules, metal oxides and hydrides, and atomic species that could be
present in hot Jupiter atmospheres: H$_2$O, CO, CH$_4$, NH$_3$,
CO$_2$, TiO, AlO, VO, FeH, TiH, CrH, Na, and K
\citep{2016SSRv..205..285M}.

Furthermore, our model considers inhomogeneous cloud coverage and
scattering hazes. Our model considers cloudy regions of the atmosphere
to consist of an opaque cloud deck at pressures larger than
P$_\text{c}$ in units of bar and scattering due to hazes above the
clouds. We employed the cloud and haze parametrization of
\citet{2017MNRAS.469.1979M} in which the haze component is included as
$\sigma=a\sigma_0(\lambda/\lambda_0)^\gamma$, where $\gamma$ is the
scattering slope, $a$ is the Rayleigh-enhancement factor, and
$\sigma_0$ is the H$_2$ Rayleigh scattering cross-section
($5.31\times10^{-31}$~m$^2$) at the reference wavelength
$\lambda_0=350$~nm. For the consideration of inhomogeneous clouds and
scattering hazes, the parameter $\bar{\phi}$ is the cloud and haze
fraction cover in the planet's atmosphere. The retrieval framework
allows for the possibility of a flat spectrum (e.g., due to weak
gaseous absorption and/or a gray homogeneous cloud cover) in the
explored parameter space. Parameter estimation and Bayesian model
comparisons are performed using MultiNest \citep{2009MNRAS.398.1601F}
through the Python interface PyMultiNest \citep{2014A&A...564A.125B}.

\begin{figure*}
  \centering
  \includegraphics[width=\textwidth]{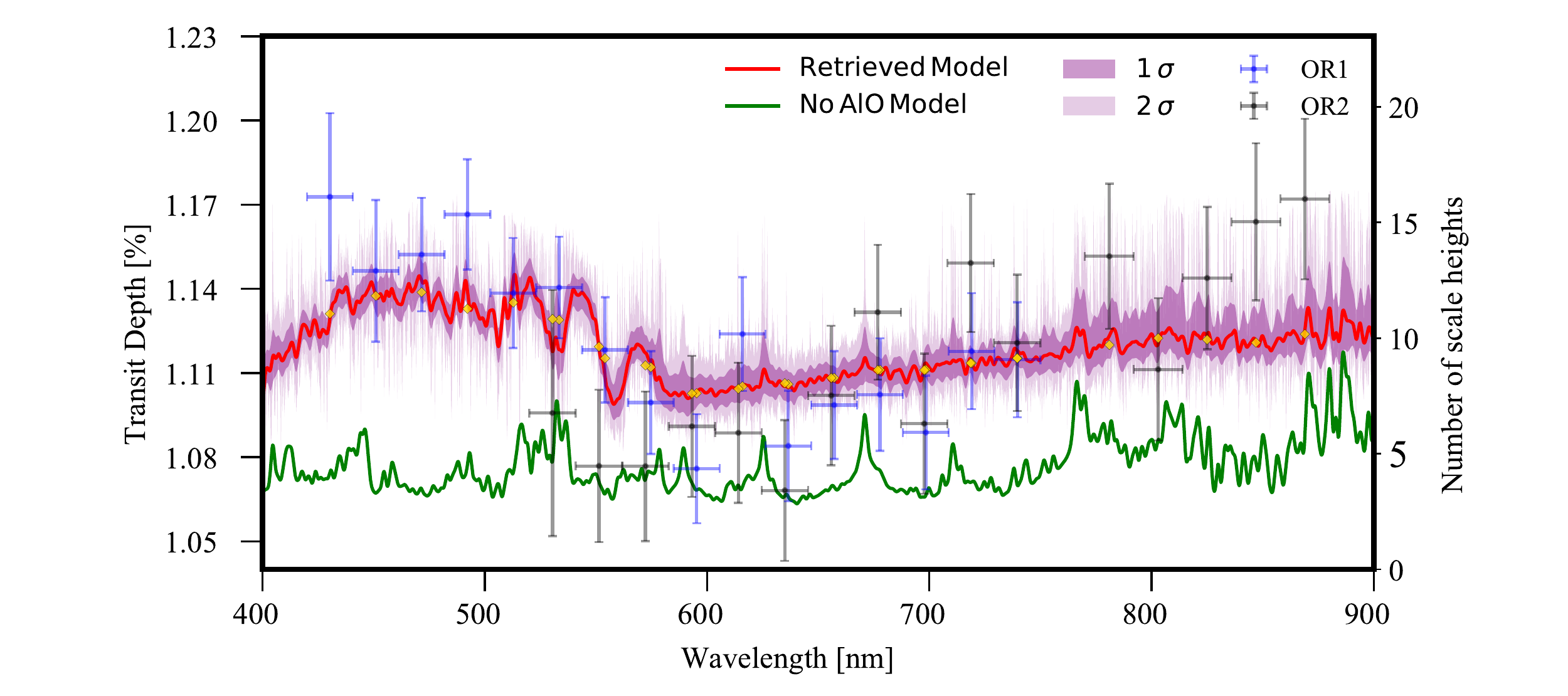}
  \caption{\label{fig:retrieval} Transmission spectrum of WASP-33b and
    retrieved models. The blue and black circles represent transit
    depths from OR1 and OR2, respectively. Horizontal lines indicate
    the size of the wavelength bin. Error bars are at the 1-$\sigma$
    level. The red curve shows the retrieved median model and the
    1-$\sigma$ and 2-$\sigma$ confidence envelopes are shown by the
    shaded regions. The gold diamonds show the binned median model at
    the same resolution as the data. The spectrum shows a feature from
    $\sim$450$-$550 nm that is explained by the model spectrum
    including AlO.}
\end{figure*}

\begin{table}[h!]
     \centering
     \caption{Bayesian model comparison detection of atmospheric compositions at the terminator of WASP-33b.}
     \label{tab:bf_retrieval}
     \begin{tabular}{lccc}
     \hline\hline\noalign{\smallskip}
     Model & Evidence   & Bayes factor & Detection \\\noalign{\smallskip}
           & $\ln(\mathcal{Z})$ & $\mathcal{B}_{0i}$ & of Ref. \\\noalign{\smallskip}
     \hline\noalign{\smallskip}
     Reference &  215.1  & Ref.   & Ref.    \\\noalign{\smallskip}
     No AlO &  211.3  & 46.7 & 3.28-$\sigma$ \\\noalign{\smallskip}
     Flat line model &  208.0  & 1232.3 & 4.18-$\sigma$ \\\noalign{\smallskip}
     No hazes/clouds &  215.8  & 0.5 & N/A  \\\noalign{\smallskip}
    \hline\noalign{\smallskip}
    \end{tabular}
\end{table}

The analysis of the transmission spectrum of WASP-33b provides initial
constraints on its atmospheric composition. Figure \ref{fig:retrieval}
shows the retrieved median fit to the observations along with the
1-$\sigma$ and 2-$\sigma$ confidence contours. In particular, we
report a possible detection of AlO at 3.3-$\sigma$ significance as
shown in Table~\ref{tab:bf_retrieval}. We retrieve a volume mixing
ratio of $\log(X_\mathrm{AlO})= -4.58 ^{+0.67}_{-0.79}$ for
AlO. Although we do not find statistically significant evidence for
the other species considered in our retrieval, the upper limits at the
99th percentile for TiO and VO are $\log(X_\mathrm{TiO})=-7.52$ and
$\log(X_\mathrm{VO})=-6.74$. Although water vapor was considered in
our models, the long-wavelength data do not show any features
corresponding to H$_2$O absorption. Models without AlO fail to explain
the features from 450-550~nm, as can be seen in
Fig.~\ref{fig:retrieval}.

We used Bayesian model comparisons to evaluate the detection
significance of AlO as shown in Table~\ref{tab:bf_retrieval}. We find
that our full 28-parameter model with AlO is preferred over a
27-parameter model without AlO at a 3.3-$\sigma$ significance. We also
investigated fits to the data with a featureless spectrum represented
by a constant transit depth, that is, a one-parameter flat line model,
using MultiNest. We find that the full 28-parameter model is preferred
over the one-parameter flat spectrum model at 4.2-$\sigma$
significance. We present the Bayesian evidence and model comparisons
in Table~\ref{tab:bf_retrieval}.

\begin{figure}
  \centering
  \includegraphics[width=.5\textwidth]{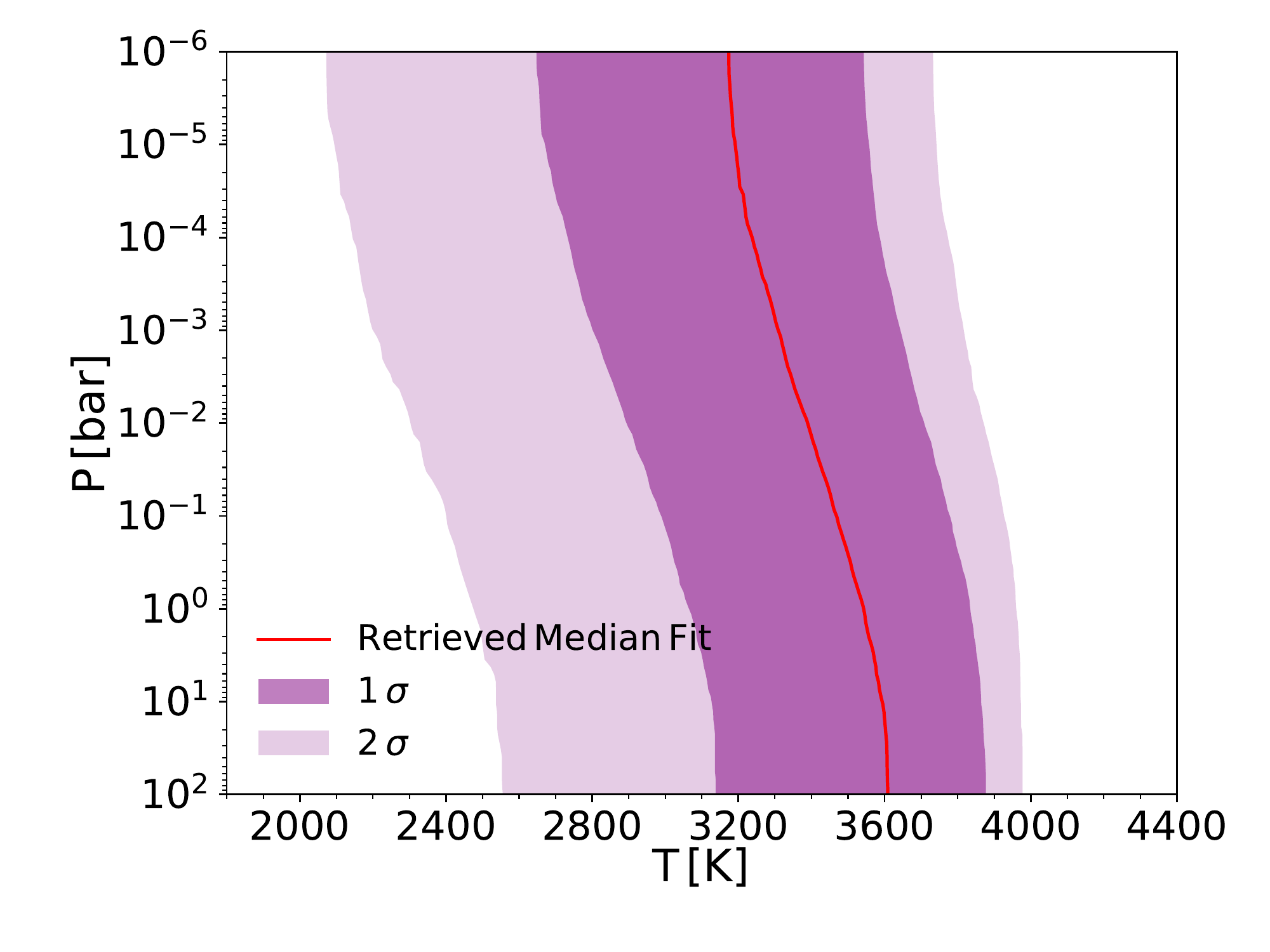}
  \caption{\label{fig:PT} Retrieved temperature structure along the
    planetary limb. The retrieved median profile is shown in red and
    the 1-$\sigma$ and 2-$\sigma$ confidence intervals are shown as
    the shaded areas.}
\end{figure}

\begin{figure*}
  \centering
  \includegraphics[width=\textwidth]{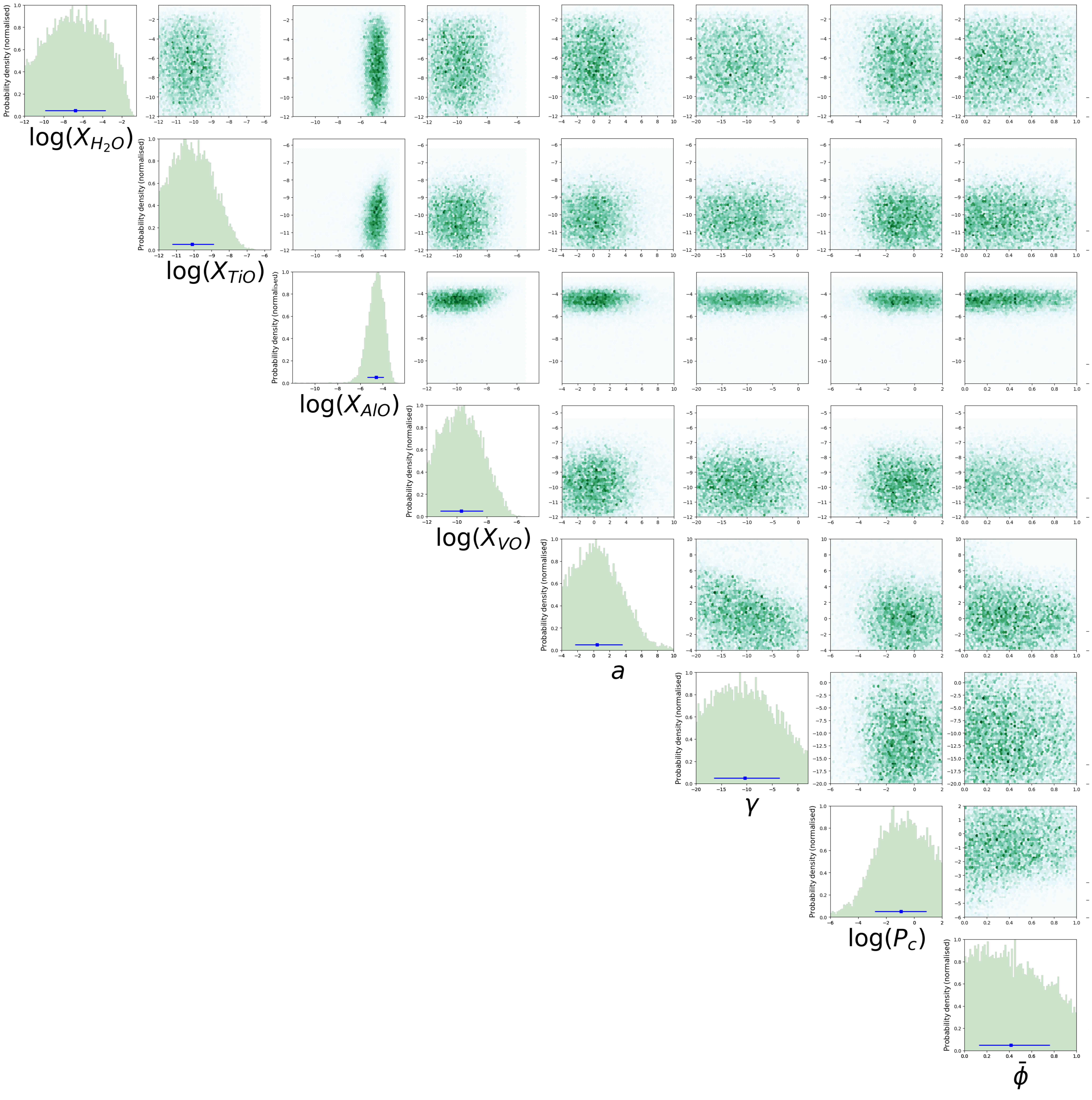}
  \caption{\label{fig:post} Marginalized posterior probability
    densities for the detected molecule (AlO) and the molecules for
    which upper bounds have been determined (TiO and VO), along with
    the cloud and haze parameters. The posterior distribution for
    water is included for reference.}
\end{figure*}

The retrieved pressure-temperature profile is shown in
Figure~\ref{fig:PT}. We obtain a relatively unconstrained profile
consistent with the equilibrium temperature of $T_{eq}\sim$2700~K,
reported in \citet{Smith2011}, within the 2-$\sigma$ region. The
retrieved median fit for the $P$-$T$ profile varies from $\sim$3200~K
at the top of the atmosphere to $\sim$3600~K at the 1-bar surface. The
retrieved $P$-$T$ profile is also consistent with the average dayside
brightness temperature of 3144 $\pm$ 114 K in the near-infrared
reported in \citet{Zhang2018}. Transmission spectra probe the
day-night terminator region, sampling temperatures of both the dayside
and nightside of the atmosphere. Furthermore, transmission spectra in
the optical probe higher regions in the atmosphere than emission
spectra. If WASP-33b has a thermal inversion
\citep[e.g.,][]{Haynes2015} it is conceivable that the upper
atmosphere probed by a transmission spectrum may be comparable in
temperature to that of the dayside photosphere reported by
\citet{Zhang2018}. Our model considers the presence of clouds and
hazes in the atmosphere of WASP-33b. However, the data does not
constrain the cloud and haze properties of the planet. The data in the
optical wavelengths lacks features indicative of a scattering
slope. We performed a retrieval test for a clear atmosphere and the
molecular abundances remained unchanged. A clear atmosphere is
consistent with studies showing that the condensation temperatures of
expected cloud- and haze-forming species are well below that of
WASP-33b \citep{2017MNRAS.471.4355P,Wakeford2017}. The posterior
distributions for the relevant parameters are shown in
Figure~\ref{fig:post}.

Our detection significance for AlO of 3.3-$\sigma$ represents a
conservative limit. Given the large number of model parameters used
for completeness in our full retrieval the evidence, and hence the
significance, is conservative. In practice, several of the 28 model
parameters do not contribute significantly in the observed visible
band. Retrieval with a model considering only the parameters that
affect the visible spectrum constitute a more meaningful measure of
the detection significance. To further test the significance of the
AlO detection we consider an additional simplified retrieval. Given
that our full retrieval was not able to constrain the cloud properties
of WASP-33b or the P-T profile, our simplified model is a clear
atmosphere with absorption from TiO and AlO only, and an isothermal
temperature profile, that is, four free parameters. The resulting
retrieval obtains a log evidence of 217.7. Using this four-parameter
model with AlO as our reference in a Bayesian model comparison, we
estimate its detection significance relative to models without AlO. We
find that the reference model is preferred over a three-parameter
model without AlO at 4.7-$\sigma$ significance. Similarly, the
reference models is preferred over a one-parameter flat line model at
4.8-$\sigma$ significance. Although this analysis suggests that our
detection of AlO is at a confidence level higher than 3.3-$\sigma$, we
still adopt the more conservative detection significance obtained
using the full 28-parameter model.

Our retrieved AlO abundance constrains the Al/H ratio in the
atmosphere. The volume mixing ratio of $\log(X_\mathrm{AlO})= -4.58
^{+0.67}_{-0.79}$ corresponds to an abundance of
$\log(\mathrm{AlO/H})= -4.81 ^{+0.67}_{-0.79}$. Assuming all the Al is
contained in AlO, our derived estimate of Al/H is consistent with a
solar value of $\log(\mathrm{Al/H})= -5.55 \pm 0.03$
\citep{2009ARA&A..47..481A}. However, Al can also be present in other
refractory species, in which case our derived estimate is a lower
limit on the true Al/H abundance in the atmosphere of WASP-33b. In
particular, the dominant Al-containing species in a solar-composition
atmosphere in equilibrium, at temperatures near 3000 K, include Al and
AlH \citep{Woitke2018}. Under such conditions, our retrieved AlO
abundance is $\sim$ 10$^3$ $\times$ higher than that predicted in
chemical equilibrium with solar abundances. Future studies may
investigate the feasibility of our retrieved abundance of AlO in
WASP-33b, possibly due to chemical disequilibrium or other
mechanisms. Future observations may also provide better constraints on
the same.

\subsection{The aluminum oxide feature}

The main spectral feature of the optical transmission spectrum
reported in this work is an increase in the planet-star radius ratio
blueward of about 560~nm. In general, an increase toward short
wavelengths can be caused by third light of an additional stellar
companion in the aperture of flux extraction. \cite{Moya2011} detected
a stellar object 2~arcsec away from \wasp, well included within our
aperture. The authors estimated that the probable physical companion
of \wasp\ should have an effective temperature of about $\sim$3000
K. As carried out in \cite{vonEssen2015}, within our wavelength region
we estimated that the M dwarf should be at least $10^3$ times dimmer
than \wasp. This would create a variability of planet-to-star radius
ratio as large as 60 ppm. This upper value is well contained within
our error bars given that the median precision of the dataset is 245
ppm. Thus, it is safe to neglect the contribution of \wasp's
companion. Also, star spots might in principle cause a spectral slope
in the transmission spectrum \citep{Sing2011,Oshagh2014}. This
possibility plays no role for \wasp\ either, since it has no or a too
thin outer convection zone, making star spots unlikely to occur. In
consequence, we have interpreted the spectral feature shortward of
560~nm as an increase in opacity in the planetary atmosphere of
\waspb.

In Section \ref{sec_retriev} we retrieved AlO as the opacity source of
the spectral feature between 450 nm and 550 nm. Spectral absorption
features by gaseous AlO have not been reported in the literature of
exoplanet observations, though atmospheric models predict its
importance in hot Jupiters (Gandhi, S., \& Madhusudhan, N., 2018,
MNRAS (submitted). AlO was described as a potential condensate forming
clouds or hazes in very hot exoplanet atmospheres
\citep[e.g.,][]{Lodders2002,Wakeford2017,2017MNRAS.471.4355P}. AlO is
difficult to observe for M or L type stellar objects, first because
the optical spectra of these low-mass stars are dominated by TiO and
VO absorption features, and second because such spectral feature at
$\sim$\,500~nm is challenging to observe for very late type stars
because of their intrinsic faintness at these wavelengths. As for the
exoplanet atmospheres, AlO is mentioned in the literature of low mass
stellar objects as condensate forming species
\citep[e.g.,][]{Burrows1999,Allard2011,Helling2017}.

\section{Conclusions}
\label{Concl} 

In this work we report GTC/OSIRIS low resolution spectroscopic
observations of \waspb\ carried out during two transits taken 18
orbits apart. Our combined observations cover wavelengths between 420
and 880 nm. Our main aim is to characterize the chemical composition
of the atmosphere of the ultra-hot Jupiter \waspb, one of the hottest
exoplanets known to date. We used our previous knowledge of the
pulsations of the host star to create a physical model that, besides
the intrinsic variability of the host star, included the transit
feature and the effects induced by our Earth's atmosphere, along with
the imperfections in the instruments collecting these data. Fitting
the chromatic transit light curves simultaneously, but the pulsations
and the planet-to-star radii ratios independently, allowed us to
contrast our results. Within the common wavelengths, our derived
transmission spectra are fully consistent within 1-$\sigma$
uncertainties. Using detailed atmospheric retrieval methods we find
that the feature observed between 450 and 550 nm can be best explained
by aluminum oxide in the planetary atmosphere at a detection
significance of 3.3-$\sigma$. We find no significant evidence for
other chemical species, but report upper limits for TiO and VO which
are indicative of subsolar abundances of these species at the
terminator region of \waspb. The obtained transmission spectrum does
not constrain the cloud and haze properties of the planet. While the
data shows strong evidence for AlO, we note that the retrieved AlO
abundance is $\sim$ 10$^3$ $\times$ higher than that predicted
assuming thermochemical equilibrium with solar elemental
abundances. More observations of the transmission spectrum of WASP-33b
in the visible, both from ground-based facilities \cite{Sedaghati2017,
  Chen2018} as well as with HST \cite{Sing2016}, could help improve
upon the present constraints.

\begin{acknowledgements}

CvE acknowledges Brandon Tingley and Tina Santl Temkiv for fruitful
discussions. Funding for the Stellar Astrophysics Centre is provided
by The Danish National Research Foundation (Grant agreement no.:
DNRF106). Based on observations made with the Gran Telescopio Canarias
(GTC), installed in the Spanish Observatorio del Roque de los
Muchachos of the Instituto de Astrofísica de Canarias, in the island
of La Palma. LW and AP are grateful to the Gates Cambridge Trust for
research support. NM acknowledges support from the UK Science and
Technology Facilities Council.

\end{acknowledgements}

\bibliographystyle{aa}
\bibliography{wasp33}

\end{document}